# Far from equilibrium maximal principle leading to matter self-organization

Piero Chiarelli

*National Council of Research of Italy, Area of Pisa, 56124 Pisa, Moruzzi 1, Italy*

*Interdepartmental Center "E.Piaggio" University of Pisa*

Phone: +39-050-315-2359
Fax: +39-050-315-2166
Email: pchiare@ifc.cnr.it.

In this work an extremal principle driving the far from equilibrium evolution of a system of structureless particles is derived by using the stochastic quantum hydrodynamic analogy. For a classical phase (i.e., the quantum correlations decay on a distance smaller than the mean inter-molecular distance) the far from equilibrium kinetic equation can be cast in the form of a Fokker-Plank equation whose phase space velocity vector maximizes the dissipation of the energy-type function, named here, stochastic free energy.
Near equilibrium the maximum stochastic free energy dissipation (SFED) is shown to be compatible with the Prigogine's principle of minimum entropy production. Moreover, in the case of elastic molecular collisions and in absence of chemical reactions, in quasi-isothermal far from equilibrium states, the theory shows that the maximum SFED reduces to the maximum free energy dissipation.
Following the tendency to reach the highest rate of SFED, the system transition to states with higher free energy can happen. Given that in incompressible ordinary fluids such an increase of free energy is almost given by a decrease of entropy, the matter self-organization becomes possible.
When chemical reactions or relevant thermal gradients are present, the theory highlights that the Sawada enunciation of maximum free energy dissipation can be violated.
The proposed model depicts the Prigogine's principle of minimum entropy production near-equilibrium and the far from equilibrium Sawada's principle of maximum energy dissipation as two complementary principia of a unique theory where the latter one is a particular case of the more general one of maximum stochastic free energy dissipation.

## INTRODUCTION

The research in the field of order generation and matter self-assembling dates back to the thirties [1-8]. Various extremal principles have been proposed for the self-organized régimes governed by classical linear and non-linear non-equilibrium thermodynamic laws, with stable stationary configurations being particularly investigated.
Nevertheless an organic understanding is still not available. In 1945 Prigogine [1,2] proposed the "Theorem of Minimum Entropy Production" which applies only to near-equilibrium stationary state. The proof offered by Prigogine is open to serious criticism [3]. Šilhavý [4] offers the opinion that the extremal principle of [near-equilibrium] thermodynamics does not have any counterpart for far from-equilibrium steady states despite many claims in the literature."
Sawada [5], in relation to the earth's atmospheric energy transport process, postulated the principle of largest amount of entropy increment per unit time. He cited the work in fluid mechanics by Malkus and Veronis [6] as having proven a principle of maximum heat current, which in turn is a maximum entropy production for a given boundary condition, but this inference is not logically ever valid.

The rate of dissipation of energy appeared for the first time in Onsager's work [7] on this subject. An extensive discussion of the possible principles of extrema of entropy production and/or of dissipation of energy is given by Grandy [8]. He finds difficulty in defining the rate of internal entropy production in the general case, showing that sometimes, for the prediction of the course of a process, the extremum of the rate of dissipation of energy may be more useful than that of the rate of entropy production.
Sawada and Suzuky [9] confirmed, both by numerical simulations and by experiments, the maximum rate of energy dissipation in electro-convective instabilities.
Nowadays, the debate about the principle of maximum free energy dissipation (MFED) and the Prigogine one's is still going on.
An alternative approach to the far from equilibrium evolution can be obtained in term of Langevin equations that in some cases describe the underlying dynamics at a continuous coarse-grained scale. The Langevin equation can be derived by using different techniques, such as the Poisson transformation [10] and the Fock space formalism [11]. Occasionally, exact formulations exist for non-linear reaction kinetics and others few problems. Alternatively, a Langevin equation can be assumed on a phenomenological point of view where it is decided *a priori* what is pertinent to the approximated dynamics. In this context it is really difficult to have a rigorous Langevin description.
The way out is to derive satisfactory Langevin equations from a microscopic model.
In the present work, by using the stochastic quantum hydrodynamic analogy (SQHA) [12-15] as the microscopic model, the classical non-equilibrium kinetics has been derived for the macro-scale limit.
The SQHA, where the structureless molecules are described by a pseudo-Gaussian wave function, allows deriving the far-from-equilibrium phase-space evolutionary criterion for classical gas and fluid phases in term of maximum dissipation of an energy-type function.

## 2. The SQHA equation of motion

The quantum hydrodynamic analogy (QHA) equations are based on the fact that the Schrödinger equation, applied to a wave function $\psi_{(q,t)} = A_{(q,t)} exp[i\frac{S_{(q,t)}}{\hbar}]$, is equivalent to the motion of a fluid owing the particle density $n_{(q,t)} = A^2_{(q,t)}$ with a velocity $\dot{q} = \frac{\nabla_q S_{(q,t)}}{m}$, governed by the equations [12-14]

$$\partial_t n_{(q,t)} + \nabla_q \bullet (n_{(q,t)} \dot{q}) = 0, \qquad (1)$$

$$\dot{q} = \nabla_p H, \qquad (2)$$

$$\dot{p} = -\nabla_q (H + V_{qu}), \qquad (3)$$

where $H$ is the Hamiltonian of the system and $V_{qu}$ is the quantum pseudo-potential that reads

$$V_{qu} = -(\frac{\hbar^2}{2m}) n^{-1/2} \nabla_q \bullet \nabla_q n^{1/2}. \qquad (4)$$

For the purpose of this paper, it is useful to observe that equations (1-3) can be derived by the following phase-space equations [15]

$$\partial_t \rho_{(q,p,t)} + \nabla \bullet (\rho_{(q,p,t)} (\dot{x}_H + \dot{x}_{qu})) = 0 \qquad (5)$$

where

$$n_{(q,t)} = \iiint \rho_{(q,p,t)} d^{3n} p. \qquad (6)$$

$$\dot{x}_H = (\nabla_p H, -\nabla_q H) \tag{7}$$

$$\dot{x}_{qu} = (0, -\nabla_q V_{qu}) \tag{8}$$

where the phase space wave function moduls (WFM) density $\rho$ has the form

$$\rho_{(q,p,t)} = \mathrm{n}_{(q,t)}\delta(p - \nabla_q S) \quad , \tag{9}$$

where

$$S = \int_{t_0}^{t} dt(\frac{p \bullet p}{2m} - V_{(q)} - V_{qu}) \, . \tag{10}$$

Due to the fact that the ensemble of solutions of equations (5-8) is wider than that one of the QHA-equations (1-3), the accessory condition

$$\dot{q} = \frac{\nabla_q S}{m} \quad , \tag{11}$$

appears in (9) [15].

The Madelung approach, as well as the Schrödinger one, are non-local and are not able to give rise to the classical limit and/or local kinetics.

Nevertheless, when fluctuations are added to the SQHA equation of motion, the resulting stochastic-QHA SQHA dynamics preserve the quantum behavior on a scale shorter than the quantum correlation length $\lambda_c$ [15]

$$\lambda_c = (\frac{\pi}{2})^{3/2} \frac{\hbar}{(2mk\Theta)^{1/2}} \tag{12}$$

In a preceding paper [15] the author has shown that in presence of vanishing small stochastic Gaussian noise, of amplitude $\Theta$, the QHA motion equation reads

$$\partial_t \rho_{(q,p,t)} = -\nabla \bullet (\rho_{(q,p,t)}(\dot{x}_H + \dot{x}_{qu(\mathrm{n}_0)})) + \eta_{(q,t,\Theta)}\delta(p - \nabla_q S), \tag{13}$$

with

$$\rho_{(q,p,t)} = \mathrm{n}_{(q,t)}\delta(p - \nabla_q S), \tag{14}$$

$$\partial_t \mathrm{n}_{0(q,t)} = -\nabla_q \bullet (\mathrm{n}_{0(q,t)} \dot{q}_{(\mathrm{n}_0)}) \tag{15}$$

where $\Theta$ is a measure of the noise amplitude and where

$$< \eta_{(q_\alpha,t)}, \eta_{(q_\beta+\lambda,t+\tau)} >= \underline{\mu} \frac{k\Theta}{\lambda_c^{\,2}} exp[-(\frac{\lambda}{\lambda_c})^2 ]\delta(\tau)\delta_{\alpha\beta} \tag{16}$$

$$S = \int_{t_0}^{t} dt(\frac{p \bullet p}{2m} - V_{(q)} - V_{qu(\mathrm{n})}) = \int_{t_0}^{t} dt(\frac{p \bullet p}{2m} - V_{(q)} - V_{qu(\mathrm{n}_0)} - I^*) \tag{17}$$

$$m\dot{q} = p = \nabla_q S = \nabla_q \{ \int_{t_0}^{t} dt(\frac{p \bullet p}{2m} - V_{(q)} - V_{qu(\mathrm{n}_0)} vI^*)\} = p_0 + \Delta p_{st} \tag{18}$$

where

$$\Delta p_{st} = -\nabla_q \{ \int_{t_0}^{t} I^* dt \}, \qquad (19)$$

where

$$I^* = V_{qu(\mathrm{n})} - V_{qu(\mathrm{n}_0)} \qquad (20)$$

### 2.1. Quantum non-locality length $\lambda_L$

In addition to the noise correlation function (12), to obtain the local form of equations (11-18) we need to evaluate the range of interaction of the quantum force $\dot{p}_{qu} = -\nabla_q V_{qu}$ in (15).

As shown in reference [15] the relevance of the quantum potential force at large distance can be evaluated by the convergence of the integral

$$\int_0^{\infty} | q^{-1} \nabla_q V_{qu} | dq < \infty \qquad (21)$$

So that the quantum potential range of interaction can be obtained as the mean weighted distance

$$\lambda_L = 2 \frac{\int_0^{\infty} | q^{-1} \frac{dV_{qu}}{dq} | dq}{\lambda_c^{-1} | \frac{dV_{qu}}{dq} |_{(q=\lambda_c)}} . \qquad (22)$$

The expression (22) generalized to a system of a large number of particles is quite complex [15]; nevertheless for the interaction of a couple of particles (e.g., mono-dimensional case, real gas or a chain of neighbors interacting atoms) the expression (22) is manageable [16].

### 2.2. Macroscopic local limiting dynamics

Given $\Delta L$ the physical length of the system, the macroscopic local dynamics is achieved for those problems that satisfy the condition

$$\lambda_c \cup \lambda_L << \Delta L .$$

From the condition $\lambda_L << \Delta L$ it follows that [15]

$$\lim_{q/\lambda_L \to \infty} -\nabla_q V_{qu(\mathrm{n}_0)} = 0 \qquad (23)$$

and the SPDE of motion acquires the form [15]

$$\partial_t \rho_{(q,p,t)} = -\nabla \bullet ( \rho_{(q,p,t)} \dot{x}_H ) + \eta_{(q,t,\Theta)} \delta( p - \nabla_q S ) \qquad (24)$$

$$\rho_{(q,p,t)} = \mathrm{n}_{(q,t)} \delta( p - \nabla_q S ), \qquad (25)$$

$$\partial_t \mathrm{n}_{(q,t)} = -\nabla_q \bullet ( \mathrm{n}_{(q,t)} \dot{q}_{cl} ) + \eta_{(q_\alpha,t,\Theta)} \qquad (26)$$

$$< \eta_{(q_\alpha,t)}, \eta_{(q_\alpha+\lambda,t+\tau)} > = \underline{\mu} \, \delta_{\alpha\beta} \frac{2k\Theta}{\lambda_c} \delta(\lambda) \delta(\tau) \qquad (27)$$

$$\dot{q} = \frac{p}{m} = \nabla_q \lim_{\Delta\Omega/\lambda_L \to \infty} \frac{\nabla_q S}{m} = \nabla_q \{ \lim_{\Delta\Omega/\lambda_L \to 0} \frac{1}{m} \int_{t_0}^{t} dt( \frac{p \bullet p}{2m} - V_{(q)} - V_{qu(\mathrm{n})} - I^*) \}$$

$$= \frac{1}{m} \nabla_q \{ \int_{t_0}^{t} dt\, ( \frac{p \bullet p}{2m} - V_{(q)} - \Delta) \} = \frac{p_{cl}}{m} + \frac{\delta p}{m} \cong \frac{p_{cl}}{m} \tag{28}$$

where $\delta p$ is a small fluctuation of momentum and

$$\dot{p}_{cl} = -\nabla_q V_{(q)}, \tag{29}$$

Being also $\lambda_c << \Delta L$, $I^*$ is a small fluctuating quantity [15].

## 3. The SQHA single-particle distribution

In the independent molecule description (that in the case of Lennard Jones potentials is possible when the range of quantum potential interaction (of order of $r_0$ [16]) is smaller than the mean intermolecular distance) between two consecutive molecular collisions, we consider the SQHA equation (1,5) for the single molecule with the noise $\eta_{(q_\alpha,t)}$

$$\partial_t \rho_{(q,p,t)} = -\nabla \bullet ( \rho_{(q,p,t)} \dot{x}_H ) + \eta_{(q,t,\Theta)} \delta( p - \nabla_q S ) \tag{30}$$

$$\partial_t \mathrm{n}_{(q,t)} = -\nabla_q \bullet ( \mathrm{n}_{(q,t)} \dot{q}_{cl} ) + \eta_{(q_\alpha,t,\Theta)} \tag{31}$$

$$< \eta_{(q_\alpha,t)}, \eta_{(q_\alpha+\lambda,t+\tau)} > = \underline{\mu}\, \delta_{\alpha\beta} \frac{2k\Theta}{\lambda_c} \delta(\lambda)\delta(\tau) \tag{32}$$

that can be re-cast in the form

$$\partial_t \mathrm{n}_{(q,t)} = -\nabla_q \bullet ( \mathrm{n}_{(q,t)} \dot{q}_{cl} + \delta\mathrm{N}_{(q,t,\Theta)} ) \tag{33}$$

$$\partial_t \rho_{(q,p,t)} = -\nabla \bullet ( \rho_{(q,p,t)} ( \dot{x}_H + \Delta \dot{x} )) \tag{34}$$

$$\Delta \dot{x} = \rho^{-1} \begin{pmatrix} \delta\mathrm{N} \delta( p - \nabla_q S ) \\ 0 \end{pmatrix} \tag{35}$$

where [17]

$$< \nabla_q \bullet \delta\mathrm{N}_{(q_\alpha,t)}, \nabla_\lambda \bullet \delta\mathrm{N}_{(q_\beta+\lambda,t+\tau)} > = \nabla_q \bullet \nabla_\lambda < \delta\mathrm{N}_{(q_\alpha,t)}, \delta\mathrm{N}_{(q_\beta+\lambda,t+\tau)} >$$

$$= < \eta_{(q_\alpha,t)}, \eta_{(q_\beta+\lambda,t+\tau)} > = \underline{\mu}\, \delta_{\alpha\beta} \frac{2k\Theta}{\lambda_c} \delta(\lambda)\delta(\tau) \tag{36}$$

Discretizing the spatial coordinates by a cell of side $\delta L$, with $\delta L > \lambda_c$, for the Markov process (32) we can write [17]

$$< \delta\mathrm{N}_{(q_\alpha,t)}, \delta\mathrm{N}_{(q_\beta+\lambda,t+\tau)} > = \mathrm{n}_{(q,t)} 2D_{(q)}\, \delta_{\alpha\beta} \delta(\lambda)\delta(\tau). \tag{37}$$

where $D_{(q)}$ is defined positive as well as $\mathrm{n}_{(q,t)}$. By comparing (36) with (37) we obtain

$$2\nabla_q \mathrm{n}_{(q,t)} D_{(q)}\, \nabla_\lambda\, ln[\delta(\lambda)] = \underline{\mu} \frac{2k\Theta}{\lambda_c} \tag{38}$$

where

$$\nabla_{\lambda} ln[\delta(\lambda)] = \nabla_{\lambda} ln \lim_{(\lambda_c \to 0)} [\frac{1}{2\lambda_c} exp[-\frac{|\lambda|}{\lambda_c}] = \nabla_{\lambda} [ \lim_{(\lambda_c \to 0)} ln[\frac{1}{2\lambda_c}] - \frac{|\lambda|}{\lambda_c}]$$

$$= \lim_{(\lambda_c / \lambda) \to 0} -\frac{1}{\lambda_c} \qquad (\lambda > 0)$$

$$= \lim_{(\lambda_c / \lambda) \to 0} \frac{1}{\lambda_c} \qquad (\lambda < 0)$$

and hence,

$$\lim_{(\lambda_c / \lambda) \to 0} \nabla_q \mathrm{n}_{(q,t)} D_{(q)} = -\vartheta_{(\lambda)} \lim_{(\lambda_c / \lambda) \to 0} \underline{\mu} k\Theta$$

where, by posing $\mathrm{n}_{(q,t)} D_{(q)} = | Aq + B |$, we obtain

$$< \delta \mathrm{N}_{(q_\alpha, t)}, \delta \mathrm{N}_{(q_\beta + \lambda, t+\tau)} >= 2 \, | Aq + B | \delta_{\alpha\beta} \delta(\lambda) \delta(\tau) .$$

that leads to

$$\lim_{(\lambda_c / \lambda) \to 0} A = -\vartheta_{(\lambda)} 2 \underline{\mu} k\Theta$$

and to

$$\lim_{(\lambda_c / \lambda) \to 0} < \delta \mathrm{N}_{(q_\alpha, t)}, \delta \mathrm{N}_{(q_\beta + \lambda, t+\tau)} >= 2 \, | -\vartheta_{(\lambda)} q \underline{\mu} k\Theta + B | \delta_{\alpha\beta} \delta(\lambda) \delta(\tau) .$$

The value of B represents the fluctuation amplitude for $\Theta = 0$ and hence it follows that B=0 and that

$$\lim_{(\lambda_c / \lambda) \to 0} < \delta \mathrm{N}_{(q_\alpha, t)}, \delta \mathrm{N}_{(q_\beta + \lambda, t+\tau)} >= \mathrm{D}_{(\Theta)} \, | q | \delta_{\alpha\beta} \delta(\lambda) \delta(\tau) \qquad (39)$$

where $\mathrm{D}_{(\Theta)} = 2 \underline{\mu} k\Theta$.

Moreover, given that the single molecule is submitted to the field of other ones $\Delta \dot{p}_{mol} = \Delta \dot{p}_{mf} + \Delta \dot{p}_{coll}$ where $\Delta \dot{p}_{mf}$ concerns the mean field of the far away molecules and $\Delta \dot{p}_{coll}$ concerns the field of the colliding molecule that comes out of the cloud and arrives at the interaction distance (for van der Waals fluids it is enough to consider just the interaction between couples of molecules [18] being three molecular collisions unlikely) we can write the SDE for the velocity as

$$\rho_{(q,t)} \dot{x} = \rho_{(q,t)} \dot{x}_H + \delta \mathrm{N}_{(q,t,\Theta)} \delta(p - \nabla_q S) + \rho_{(q,t)} \Delta \dot{x}_{coll} + \rho_{(q,t)} \Delta \dot{x}_{mf} \qquad (40)$$

where

$$\Delta \dot{x}_{coll} = (0, \Delta \dot{p}_{coll}) . \qquad (42)$$

$$\dot{\Delta x}_{mf} = (0, -\frac{\partial \overline{V}}{\partial q}) = (0, \Delta \dot{p}_{mf}). \quad (42)$$

where $\overline{V}$ is the mean-field potential of the cloud of molecules leading to the "mean field Hamiltonian"

$$\overline{H} = H + \overline{V} \quad (41)$$

Moreover, given that $\delta N_{(q,t,\Theta)}$ and $\Delta \dot{p}_{coll}$ are independent (de-coupled) owing very a different time scale (the zero correlation time for the $\Theta$ –fluctuations and the molecular collision time $\tau$ for $\Delta \dot{p}_{coll}$) the SDE (40) mediated over the zero mean $\Theta$ –fluctuations reads

$$< \dot{x} > = < \dot{x}_{\overline{H}} > + < \Delta \dot{x}_{coll} > \quad (43)$$

where

$$\dot{x}_{\overline{H}} = (\frac{\partial \overline{H}}{\partial p}, -\frac{\partial \overline{H}}{\partial q}) \quad (44)$$

owing the $\Theta$-driven noise $\delta N_{(q,t,\Theta)}$ a zero-mean.

Moreover, since the molecular dynamics in a L-J gas phase is highly chaotic, it is possible to assume that molecular collisions are not correlated and hence $< \Delta \dot{x}_{coll} >$ can be approximated as a white noise

$$< \Delta \dot{x}_{coll} > = D^{\frac{1}{2}} \xi_{(t)} \quad (45)$$

leading to the Fokker-Plank equation that written in a phase space conservation equation that reads

$$\partial_t \rho^* + \nabla \bullet (\rho^* (< \dot{x}_{\overline{H}} > + < \dot{x}_s >)) = 0 \quad (46)$$

$$< \dot{x}_s > = -(\nabla \bullet \underline{\underline{D}} + \underline{\underline{D}} \bullet \nabla \ln[\rho^*]) \quad (47)$$

where

$$\underline{\underline{D}} \equiv D_{ij} = \frac{1}{2} \tau^{-2} \int_t^{t+\tau} < \Delta \dot{x}_{coll} >_i < \Delta \dot{x}_{coll} >_j dt \quad (48)$$

Here, we derive the statistical probability density distribution $\rho^*$ in term of the single-particle SQHA equation of motion instead that by the particle point density distribution (derived by the classical motion of molecules).

Given a non-linear classic system (no quantum correlations on the scale of molecular mean distance) [19] and hence ergodic, the phase space means $\rho^* = \frac{\Delta N_\Omega}{\Delta \Omega}$, where $\Delta N_\Omega$ is the number of molecules in the phase space domain $\Delta \Omega_{(q,p)}$, coincides with the time-means.

In Ref. [19] is detailed how the SQHA "dynamical distribution" $\rho$ leads to the statistical distribution $\rho^S$.

By using the definition $\rho^S$ (see Appendix A in Ref. [19]) for the single molecule statistical distribution, we have

$$\rho^{s}{}_{1(q_j,p_h)} = \frac{\Delta n_{1hj}}{\Delta q^{3k}\Delta p^{3k}}$$

$$= \Delta q^{-3}\Delta p^{-3}\sum_i \int_{(j-1)\Delta q_1}^{j\Delta q_1} .... \int_{(j-1)\Delta q_3}^{j\Delta q_3} \left( \int_{(h-1)\Delta p_1}^{h\Delta p_1} .... \int_{(h-1)\Delta p_3}^{h\Delta p_3} P_{i(1)}(\rho) d^3p \right) d^3q \qquad (49)$$

where the operator $P_{i(1)}$ reads

$$P_{i(1)} = \int_{-\infty}^{+\infty} .... \int_{-\infty}^{+\infty} d^3q_1 d^3q_{j\neq i} d^3q_{3n} \int_{-\infty}^{+\infty} .... \int_{-\infty}^{+\infty} d^3p_1 d^3p_{j\neq i} d^3p_{3n} \qquad (50)$$

Equation (50) for real gas (and Markovian van der Waals fluids), using the independent particle description with $\rho \cong \prod_i \rho_{(i)}$, leads to the explicit the connection with $\rho^*$ that reads

$$\rho^{s}{}_{1(q_j,p_h)} = \Delta q^{-3}\Delta p^{-3}\sum_i \int_{(j-1)\Delta q_1}^{j\Delta q_1} .... \int_{(j-1)\Delta q_3}^{j\Delta q_3} \left( \int_{(h-1)\Delta p_1}^{h\Delta p_1} .... \int_{(h-1)\Delta p_3}^{h\Delta p_3} P_{i(1)}(\rho) d^3p \right) d^3q$$

$$= \frac{\sum_i \int_{\Delta\Omega_{(q,p)}} \rho_{(i)} d^3q d^3p}{\Delta\Omega} = \frac{\Delta N_{\Omega}}{\Delta\Omega} = \rho^* \qquad . (51)$$

Thence, (46-47) read

$$\lim_{\Delta L >> \lambda_c, \lambda_L} \partial_t \rho^s + \nabla \bullet (\rho^s (<\dot{x}_{\overline{H}}> + <\dot{x}_s>)) = 0 \qquad (52)$$

$$\lim_{\Delta L >> \lambda_c, \lambda_L} \rho^s <\dot{x}_s> = -\nabla \bullet (\underline{\underline{D}} \rho^s) \qquad (53)$$

The equality $\rho^* = \rho^s$ follows by the absence of quantum correlation between particles and by the ergodicity, that for systems of a huge number of particle in the SQHA model is warranted by the non-linearity that is necessary to the establishing of the classical behavior [15-16].

For a (classical) gas phase made up of structureless point-like particles interacting by central symmetric potential that do not undergo to chemical reactions (particles do not have bounded states (e.g., Lennard-Jones potential with small well, compared do the mean energy of particles) so that molecules with internal structure are not created) (53) can be further simplified by excluding the cross-correlations concerning different co-ordinates components, namely

$$D_{ij} = D_{(i)}\delta_{ij} = \begin{pmatrix} 3D_q & 0 \\ 0 & 3D_p \end{pmatrix} = D \begin{pmatrix} 3I_q & 0 \\ 0 & 3I_p \end{pmatrix} \qquad (54)$$

Disregarding the out of diagonal terms of the diffusion matrix $D_{ij}$, (53) reads

$$\lim_{\Delta L >> \lambda_c, \lambda_L} \rho^s <\dot{x}_s> = -\nabla D_{(i)} \rho^s \qquad (55)$$

Equation (55) is nothing less than the Fokker-Plank form of Maxwell equation. The difference with the Boltzmann kinetic equation is that it cannot give any information about the form of the (phase space) diffusion coefficient $D$. In order to obtain from (55) a closed kinetic equation, additional information about the diffusion coefficient must be introduced. In the local equilibrium limit this can be done by the semi-empirical assumption of linear relation between flows and fluxes (the momentum components of $D_p$ are related to the physical viscous constants).

On the other hand, it must be observed that even if the Boltzmann kinetic equation is able to lead by itself to the explicit form of the diffusion coefficient near equilibrium, (55) is of general validity and it holds even far from equilibrium.

## 4. The kinetic equation for the classical phase

### 4.1 The mean phase space molecular volume of WFM

In order to catch additional information from (55) observe that (for gasses and Markovian fluids) the SQHA approach shows two competitive dynamics: (1) the enlargements of the molecular WFM (given by (31)) between two consecutive collisions) (2) The diffusion of the molecules, in term of their mean position, as a consequence of the molecular collisions that cause the WFM collapse [16].

As consequence of free expansions and collapses, the pseudo-Gaussian molecular WFM in the phase space cell $\Delta\Omega$ will occupy the volume $< \Delta V_m >$ that we pose

$$\lim_{\Delta L >> \lambda_c, \lambda_L} < \Delta V_m > = h^3 \exp[-\phi] \tag{56}$$

where $< \Delta V_m >$ reads:

$$< \Delta V_m > = \frac{\sum\limits_{i \in \Delta\Omega} \left( \int\limits_{\Delta\Omega_{(q,p)}} \rho_{(i)} (x_{(i)} - < x_{(i)} >)^2 d^3q d^3p \right)^{\frac{1}{2}}}{\sum\limits_{i \in \Delta\Omega} \int\limits_{\Delta\Omega_{(q,p)}} \rho_{(i)} d^3q d^3p} \tag{57}$$

$$< x_{(i)} > = \frac{\int\limits_{\Delta\Omega_{(q,p)}} \rho_{(i)} x_{(i)} d^3q d^3p}{\int\limits_{\Delta\Omega_{(q,p)}} \rho_{(i)} d^3q d^3p} \tag{58}$$

where $< x_{(i)} > = \begin{pmatrix} q_{(i)} \\ p_{(i)} \end{pmatrix}$. Given the mean WFM volume (WFMV) per molecule $< \Delta V_m >$ to be a fraction "α" of the phase space volume available per molecule $\frac{\Delta\Omega}{\Delta N_\Omega}$ we can pose

$$< \Delta V_m > = \alpha \frac{\Delta\Omega}{\Delta N_\Omega} \tag{59}$$

where $\Delta N_\Omega$ is the number of molecules in $\Delta\Omega$.

In the case of stationary states, $\alpha$ comes from the balance of two contributes: (1) Proportional to the diffusive enlargement of WFM (generated by the diffusion coefficient $\mathrm{D}_{(\Theta)} = 2\underline{\mu} k\Theta$) and (2) proportional to the collision time (inversely proportional to the diffusion coefficient $D$). Therefore, we can formally write

$$\alpha = \alpha' \frac{\mathrm{D}_{(\Theta)}}{D} = \frac{D^*}{D} \tag{60}$$

where $D^*$ can be thought as an effective diffusion coefficient comprehending $\alpha'$.
As shown in appendix B, the variation of the absolute value of the constant $\alpha'$ leads to the re-definition (by a constant) of the free energy at thermodynamic equilibrium. Moreover, by using the definition of $\rho^s$

$$\rho^s = \frac{\Delta N_\Omega}{\Delta \Omega} \tag{61}$$

From (59), for stationary states we have

$$\lim_{\Delta L >> \lambda_c, \lambda_L} < \Delta V_m >= h^3 exp[-\phi] = \frac{D^*}{D} {\rho^s}^{-1} \tag{62}$$

$$\lim_{\Delta L >> \lambda_c, \lambda_L} \rho^s = h^{-3} \frac{D^*}{D} exp[\phi] \tag{63}$$

$$\lim_{\Delta L >> \lambda_c, \lambda_L} < \dot{x}_s >= -D\{\nabla \phi + \nabla ln[D^*]\} \tag{64}$$

where (55) has been used.
Given that at thermodynamic equilibrium (in absence of external fields) there is the transnational invariance in the phase space, it holds $\nabla \phi = 0$, as well as $D^*$ constant and hence $\nabla ln[D^*] = 0$.
The same result will apply if the $\Theta$ -fluctuations and the thermal ones are decoupled. In this case, we could assume $\Theta$ =constant independently by the thermodynamic conditions and hence $\nabla ln[D^*] = 0$ even out of equilibrium.
Actually, the coupling between the amplitude $\Theta$ of the WFM energy fluctuations and the thermal ones must be considered.
Both the WFM fluctuations and the thermal ones are the consequence of large environments: 1) The vacuum energy fluctuations for the WFM [15], 2) the thermal energy of the environmental matter for the particles.
The fact that the coupling between matter and vacuum must be considered becomes clear if we try to reach the zero temperature T by step of equilibrium. In this case, we will end with a no fluctuating state of the system (i.e., a deterministic quantum state) and this will be not possible in presence of vacuum fluctuations ($\Theta \neq 0$). This can be easily verified by calculating the quantum mean energy fluctuation by starting from the SQHA model.
Assuming a coupling between matter and vacuum, hence, as we diminish the thermal energy fluctuation we will also lower the vacuum ones.
This also constitutes everyday evidence since open quantum phenomena are elicited by the temperature lowering.
In the case of sufficiently weak radiative coupling we can write

$$\nabla ln[D^*] = \underline{\underline{A}} \bullet \nabla \phi + \underline{\underline{\underline{B}}} \bullet \nabla \phi \nabla \phi + O(\nabla \phi^3) \tag{65}$$

In the case of structureless punt-like particles, classically interacting (i.e., the classical gas phase) with central-symmetric potential, the direction of variation of $D^*$ must be aligned with $\nabla \phi$ leading to $\underline{\underline{A}} = A\delta_{ij}$ and $\underline{\underline{\underline{B}}} = 0$ and, hence, to

$$\nabla ln[D^*] = A \nabla \phi + O(\nabla \phi^3) = (A + O(\nabla \phi^2)) \nabla \phi \tag{66}$$

from where it follows that

$$\lim_{\Delta L >> \lambda_c, \lambda_L} < \dot{x}_s >= -D \nabla \phi (1 + A + O(\nabla \phi^2)) \tag{67}$$

that introduced into the FPE (52-53) leads to the kinetic equation

$$\partial_t \rho^s + \nabla \bullet \rho^s < \overset{\bullet}{x}_{\overline{H}} >= \nabla \bullet \rho^s D \nabla \phi (1 + A + O(\nabla \phi^2)) \qquad (68)$$

Moreover, given that for structureless particles undergoing elastic collisions (e.g., no chemical reactions) in absence of external field, H is conserved quantity, so that it holds $\nabla \bullet < \overset{\bullet}{x}_{\overline{H}} >= 0$ (the average "< >" is done on the Θ-fluctuation of each single molecule), equation (68) for stationary states leads to

$$\partial_t \rho^s + (< \overset{\bullet}{x}_{\overline{H}} > + < \overset{\bullet}{x}_s >) \bullet \nabla \rho^s = \rho^s \nabla \bullet D \nabla \phi (1 + A + O(\nabla \phi^2)) = \frac{d\rho^s}{dt} \qquad (69)$$

From (69) it immediately follows that the thermodynamic equilibrium (i.e., $\frac{d\rho^s}{dt} = 0$ see appendix B in Ref. [19]) is obtained for $\nabla \phi = 0$ given that

$$\nabla \phi = 0 \Rightarrow \frac{d\rho^s}{dt} = 0 \qquad (70)$$

(the above results hold in absence of external potential field V, otherwise the re-definition of $\phi' = \phi + V^{ext}$ leads to the same result)

### 4.2 Far from equilibrium relaxation and maximum stochastic dissipation in stationary states

Even if the $\phi$-function is well defined far from equilibrium, the kinetic equations (68,69) without the initial and boundary condition of an assigned problem is just a symbolic equation. Nevertheless, the existence of the $\phi$-function allows the definition of a formal criterion of evolution.

By writing the irreversible phase space velocity field as follows

$$\lim_{\Delta L >> \lambda_c, \lambda_L} < \overset{\bullet}{x}_s >= -D \nabla \phi (1 + A + O(\nabla \phi^2)) \qquad (71)$$

an evolutionary principle along the relaxation pathway can be formulated in terms of dissipation of the $\phi$-function (named here *normalized hydrodynamic free energy* (NHFE) since at equilibrium it converges to the free energy normalized to kT (see appendix B in Ref. [19]).

Given that, the total differential of the *normalized hydrodynamic free-energy* $\phi$ can be written as a sum of two terms, such as:

$$\frac{d\phi}{dt} = \frac{\partial \phi}{\partial t} + < \overset{\bullet}{x} > \bullet \nabla \phi = \frac{\partial \phi}{\partial t} + < \overset{\bullet}{x}_{\overline{H}} > \bullet \nabla \phi + < \overset{\bullet}{x}_s > \bullet \nabla \phi = \frac{d_H \phi}{dt} + \frac{d_s \phi}{dt} \qquad (72)$$

where we name

$$d_H \phi = \lim_{\Delta L_L >> \lambda_c, \lambda_L} \{ \frac{\partial \phi}{\partial t} + < \overset{\bullet}{x}_{\overline{H}} > \bullet \nabla \phi \} \delta t \qquad (73)$$

"*dynamic differential*" and

$$d_s \phi = \lim_{\Delta L >> \lambda_c, \lambda_L} [ < \overset{\bullet}{x}_s > \bullet \nabla \phi ] \delta t \qquad (74)$$

"*stochastic differential*".

Under the range of validity of equation (71) (i.e., structureless punt-like particles, interacting by L-J central symmetric potential that do not undergo to chemical reactions) the stochastic velocity vector evolves through a pathway that follows the $\phi$-function negative gradient so that

$$\frac{d_s\phi}{\delta t} \text{ is minimum with respect the choice of } <\overset{\bullet}{x}_s> \tag{75}$$

and $\frac{d_s\phi}{\delta t} < 0$ since $<\overset{\bullet}{x}_s>$ is anti-parallel to $\nabla\phi$.

Sometime, some authors speak in term of energy dissipation, in this case we have

$$-\frac{d_s\phi}{\delta t} = \left|\frac{d_s\phi}{\delta t}\right| \text{ is maximum with respect the choice of } <\overset{\bullet}{x}_s> \tag{76}$$

## 5. Stability and maximum stochastic dissipation

In order to elucidate the significance of the criterion given by (76), we analyze it the spatial kinetics far and near equilibrium.

### 5.1 Spatial kinetic equations

By using a well known method [21] we transform the motion equation (68) into a spatial one over a finite volume V.

Given a quantity per particle

$$\underline{\Upsilon} = \frac{\int_{-\infty}^{+\infty}\int_{-\infty}^{+\infty}\int_{-\infty}^{+\infty} \rho^s \Upsilon d^3 p}{\int_{-\infty}^{+\infty}\int_{-\infty}^{+\infty}\int_{-\infty}^{+\infty} \rho^s d^3 p} \tag{77}$$

its spatial density:

$$n\underline{\Upsilon} = \int_{-\infty}^{+\infty}\int_{-\infty}^{+\infty}\int_{-\infty}^{+\infty} \rho^s \Upsilon d^3 p \tag{78}$$

and its first moment

$$n\underline{\Upsilon \overset{\bullet}{q}} = \int_{-\infty}^{+\infty}\int_{-\infty}^{+\infty}\int_{-\infty}^{+\infty} \rho^s \Upsilon <\overset{\bullet}{q}> d^3 p \tag{79}$$

by using the motion equation (68) it is possible to obtain the spatial differential equation:

$$\begin{aligned} \partial_t n\underline{\Upsilon} + \nabla \bullet n\underline{\Upsilon \overset{\bullet}{q}} - \int_{-\infty}^{+\infty}\int_{-\infty}^{+\infty}\int_{-\infty}^{+\infty} \rho^s \{\partial_t \Upsilon + <\overset{\bullet}{x}_{<H>}> \bullet \nabla \Upsilon\} d^3 p \\ = \int_{-\infty}^{+\infty}\int_{-\infty}^{+\infty}\int_{-\infty}^{+\infty} \Upsilon\{\nabla \bullet \rho^s D \nabla\phi(1 + A + O(\nabla\phi^2))\} d^3 p \end{aligned} \tag{80}$$

That by choosing

$$\Upsilon = kT\phi, \tag{81}$$

where T is the “mechanical” temperature defined as

$$T = \gamma \frac{<E_{cin} + E_{pot}>}{k} = \gamma\left(\frac{\frac{<p_i><p_i>}{2m} + <\overline{V}_i>}{k}\right), \tag{82}$$

where $\gamma$ is defined at thermodynamic equilibrium.

After some manipulations (see appendix C in Ref. [19]) for a system at thermal equilibrium (i.e., small thermal gradients) but far from equilibrium in terms of concentrations and mechanical parameters, at constant volume, we obtain

$$\frac{d\Phi}{dt} - \frac{d\Phi_{sup}}{dt} - \frac{d(E_{cin} + E_{int})}{dt} + \frac{dTS^{s}{}_{vol}}{dt} = -\iiint_{V} \int_{-\infty}^{+\infty}\int_{-\infty}^{+\infty}\int_{-\infty}^{+\infty} kT \frac{(\phi - 1)}{\phi} \rho^{s} (\frac{d_{s}\phi}{dt}) d^{3}p \; d^{3}q + \Delta_{0} + \Delta_{1} \tag{83}$$

where $\Delta_0$ represents the "source" term

$$\Delta_{0} = \{ \int_{-\infty}^{+\infty}\int_{-\infty}^{+\infty}\int_{-\infty}^{+\infty} \phi^{2} (\nabla \bullet \nabla \phi)(1 + A + O(\nabla \phi^{2})) d^{3}p \}, \tag{84}$$

and $\Delta_1$ the out of equilibrium contribution

$$\begin{aligned}\Delta_{1} = & \int_{-\infty}^{+\infty}\int_{-\infty}^{+\infty}\int_{-\infty}^{+\infty} \rho^{s} \{ \partial_{t} \Delta \Upsilon + < \dot{x}_{\overline{H}} > \bullet \nabla(\Delta \Upsilon) \} d^{3}p \\ & + \int_{-\infty}^{+\infty}\int_{-\infty}^{+\infty}\int_{-\infty}^{+\infty} \rho^{s} \{ T \partial_{t} \Delta S + < \dot{x}_{\overline{H}} > \bullet T \nabla \Delta S \} d^{3}p\end{aligned} \; . \tag{85}$$

(where $\Delta S = S^{s} - S$ and $\Delta \Upsilon = \Upsilon^{s} - \Upsilon^{eq}$ are the differences with the (local) equilibrium entropy $S$ and free energy $\Upsilon^{eq}$, respectively (see appendix C in Ref. [19]), where

$$\frac{d\Phi_{sup}}{dt} = \oint\!\!\oint n \underline{\Upsilon \dot{q}} \bullet d\sigma \tag{86}$$

where $d\sigma$ is a vector perpendicular to the infinitesimal element of the boundary surface, and where

$$\frac{d\Phi}{dt} = \iiint_{V} \frac{\partial (n \underline{\Upsilon})}{\partial t} d^{3}q \tag{87}$$

$$\frac{dTS^{s}{}_{vol}}{dt} = \frac{\gamma}{k} \iiint_{V} \int_{-\infty}^{+\infty}\int_{-\infty}^{+\infty}\int_{-\infty}^{+\infty} \rho^{s} S^{s} F \bullet < \dot{q} > d^{3}p \, d^{3}q = \frac{\gamma}{k} \iiint_{V} F \bullet \underline{S^{s} \dot{q}} \; d^{3}q \tag{88}$$

where

$$S^{s} = -k \ln \rho^{s} . \tag{89}$$

$$\underline{S^{s} \dot{q}} = \int_{-\infty}^{+\infty}\int_{-\infty}^{+\infty}\int_{-\infty}^{+\infty} \rho^{s} S^{s} < \dot{q} > d^{3}p \tag{90}$$

and where for potentials that are not function of momenta, $F = < \dot{p} >_{(q)}$ has been brought out of the integral in (88).
Moreover, $E_{int}$ and $E_{cin}$ are the internal energy and the macroscopic kinetic energy of the system, respectively.
Since for stationary states in quasi-isothermal condition at constant volume (fixed wall) it holds

$$\frac{d\Phi}{dt} = 0 \text{, and } \frac{dE_{int}}{dt} = 0 \text{, } \frac{dE_{cin}}{dt} = 0 \text{,} \tag{91}$$

it follows that

$$\frac{d\Phi_{sup}}{dt} = \frac{d(E_{sup} + \Delta E_{sup})}{dt} - \frac{dTS^{s}{}_{sup}}{dt} = -\frac{dTS^{s}{}_{sup}}{dt}, \tag{92}$$

$$\frac{dTS^{s}{}_{vol}}{dt} = 0 \tag{93}$$

and hence from (75) that

$$\frac{dTS_{sup}}{dt} + \frac{dT\Delta S_{sup}}{dt} - \Delta_0 - \Delta_1 = -\iiint_V \int_{-\infty}^{+\infty}\int_{-\infty}^{+\infty}\int_{-\infty}^{+\infty} kT \frac{(\phi - 1)}{\phi} \rho^{s} \left(\frac{d_s \phi}{dt}\right) d^3 p \; d^3 q \tag{94}$$

## 6. Maximum hydrodynamic free energy dissipation in stationary far from equilibrium quasi-isothermal states

The importance of stationary quasi-isothermal states far from equilibrium comes from the fact that living systems always operate in such a condition.
If we consider the overall system (environment plus system in Fig. 1) sometime the energetic reservoir is able to maintain the system (even for a long laboratory time scale) in a stationary state even the global system (system plus reservoirs) is relaxing toward the global equilibrium.

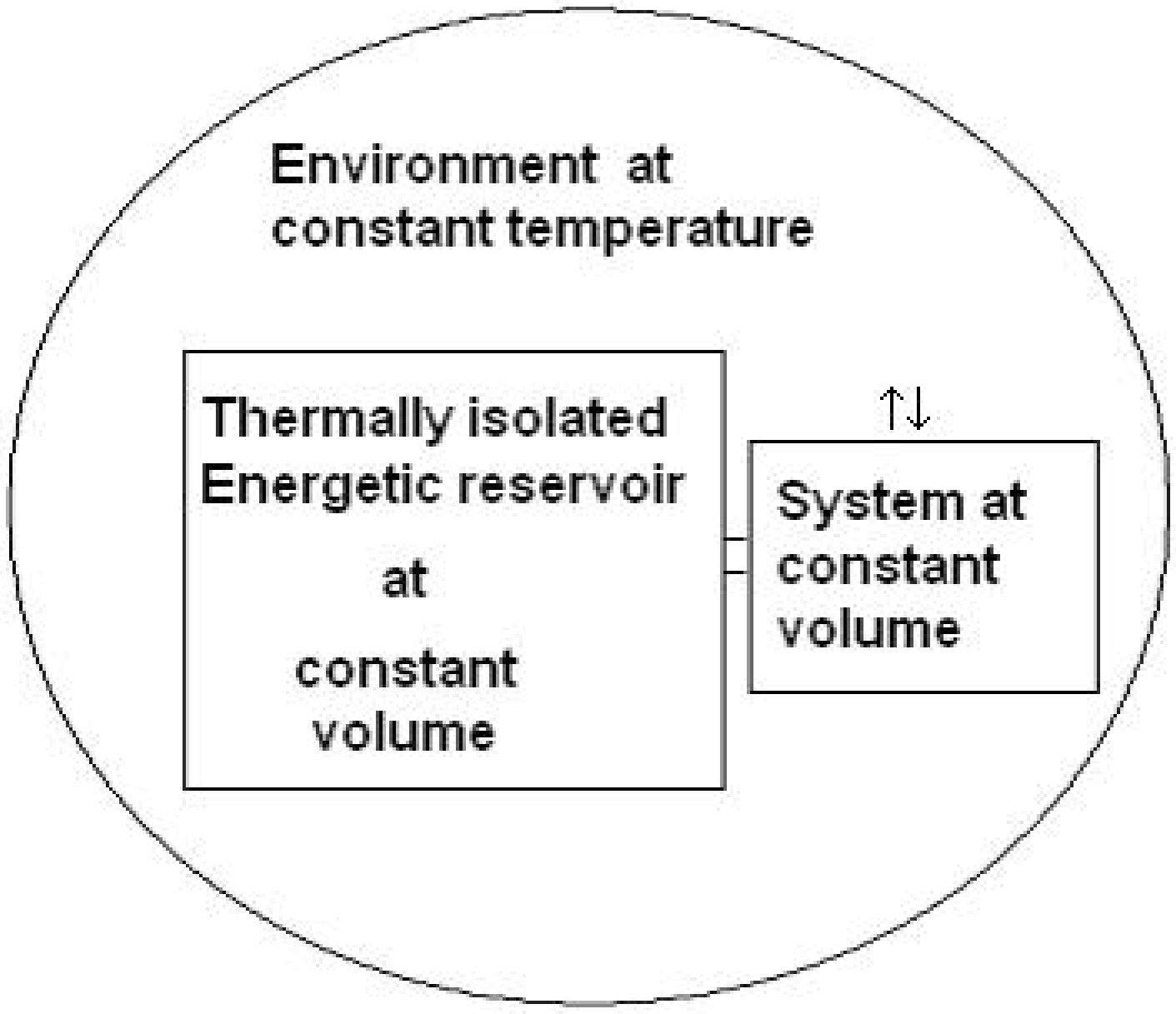

Fig.1. Schematic view of the system and the energetic reservoir and their interaction with the environment.

We consider the case where the system is driven in an instability where may exist many steady states (on time scale of relaxation of local fluctuations).
On much larger time scale, we assume that very large fluctuations can happen (i.e., the system can be driven very far from the preceding state)) and the system can make transitions among the metastable states.
For stationary states in quasi-isothermal condition we observe that the condition to be far from equilibrium consists in being far from mechanical and chemical (concentrations) equilibrium condition.

Moreover, by the quasi-isothermal condition we can infer that the local thermal equilibrium exists even if the local domains $\Delta\Omega_q{}^s$ are far from mechanical chemical equilibrium and hence and that

$$\Delta_0 \cong 0, \tag{96}$$

$$\Delta S = S^s - S \cong 0, \tag{97}$$

$$\Delta \Upsilon = \Upsilon^s - \Upsilon^{eq} \cong 0 \quad (97) \tag{98}$$

and hence that $\Delta_1 \cong 0$. So that (98) reads

$$\frac{dTS_{sup}}{dt} \cong -\iiint_V \{ \int_{-\infty}^{+\infty}\int_{-\infty}^{+\infty}\int_{-\infty}^{+\infty} kT\frac{(\phi-1)}{\phi}\rho^s(\frac{d_s\phi}{dt})d^3p\} d^3q \tag{99}$$

Moreover, assuming that the system plus the energetic reservoir are at constant volume (see Fig.1) (Without loss of generality, we can assume the energetic reservoirs are much bigger than the system and that they work on it in a reversible manner (do not dissipate energy in heat) and do not lose heat toward the environment). In this case the decrease of reservoirs free energy (no heat and energy is transferred to the environment because is thermally isolated with fixed walls) is equal to the free energy transferred on the system by means of superficial and volume forces).

Moreover, assuming that the (free) energy transferred to the system $-\frac{dE_{res}}{dt}$ is then dissipated in heat $\frac{dTS_{sources}}{dt}$ that, on its turn, is reversibly transferred to the environment through the surface $\frac{dTS_{sup}}{dt} = \frac{dQ_{sup}}{dt}$ at constant temperature (defined positive outgoing), for the energy conservation it follows that

$$\frac{dE_{res}}{dt} = -\frac{dTS_{sources}}{dt} = -\frac{dTS_{sup}}{dt} = -\frac{dQ_{sup}}{dt} \tag{100}$$

where by (96) we have

$$\frac{dTS_{sup}}{dt} = \frac{dTS^s{}_{sup}}{dt} \tag{101}$$

and, finally, that

$$-\frac{dE_{res}}{dt} = \frac{dTS_{sup}}{dt} = -\iiint_V \{ \int_{-\infty}^{+\infty}\int_{-\infty}^{+\infty}\int_{-\infty}^{+\infty} kT\frac{(\phi-1)}{\phi}\rho^s(\frac{d_s\phi}{dt})d^3p\} d^3q \tag{102}$$

that for a stationary state far from equilibrium is maximum with respect the variations of $<\dot{x}_s>$.

Therefore, for a classical phase of molecules undergoing to elastic collisions and without chemical reactions at quasi-isothermal condition, the system finds the stationary condition by maximizing its free energy dissipation.

Considering the composite system made up by the system plus the energetic reservoir (at constant volume and number of molecules) under the condition that thermostat is large and at equilibrium (both energetic reservoir and environment with relaxation constants much shorter than the system one, Sawada [9] has shown and experimentally measured in the electro-convective instability that the when the steady state configuration is achieved the system reaches the maximum of free energy dissipation (and that the more stable state among similar metastable ones owns the highest free energy dissipation).

Moreover, considering (stationary states) for system far from equilibrium with small thermal gradient (i.e., at quasi-constant temperature) where $\frac{dE_{res}}{dt}$ cannot be controlled or known as for the atmosphere turbulence (102) reads

$$\frac{dTS_{sup}}{dt} \cong \frac{TdS_{sup}}{dt} = -\iiint_V \{ \int_{-\infty}^{+\infty}\int_{-\infty}^{+\infty}\int_{-\infty}^{+\infty} kT \frac{(\phi-1)}{\phi} \rho^s (\frac{d_s\phi}{dt}) d^3p \} d^3q \qquad (103)$$

so that we also retrieve the principle of maximum heat transfer $\frac{TdS_{sup}}{dt}$ of Malkus and Veronis [6].

Here, the difference with the Sawada and the Malkus and Veronis enunciation is that (102) is not of general value but it holds only in the case of quasi-isothermal conditions for an ordinary real gas and its fluid phase made of structureless molecules (e.g., rigid-sphere) sustaining elastic collisions and not undergoing to chemical reactions.

Actually, from general point of view form (111) we see that

$$-\frac{dE_{res}}{dt} - \Delta + \frac{dT\Delta S_{sup}}{dt} = -\iiint_V \{ \int_{-\infty}^{+\infty}\int_{-\infty}^{+\infty}\int_{-\infty}^{+\infty} kT \frac{(\phi-1)}{\phi} \rho^s (\frac{d_s\phi}{dt}) d^3p \} d^3q \qquad (104)$$

is maximum with respect the variation of $< \dot{x}_s >$ so that the extremal condition of maximum free energy dissipation $-\frac{dE_{res}}{dt}$ is not as general as the right side of (104) since both the terms $\Delta$ and $\frac{dT\Delta S_{sup}}{dt}$ are not generally null.

## 6.1 Minimum entropy production in stationary states at local-equilibrium

By introducing (101) in (99) we obtain

$$-\iiint_V \{ \int_{-\infty}^{+\infty}\int_{-\infty}^{+\infty}\int_{-\infty}^{+\infty} kT \frac{(\phi-1)}{\phi} \rho^s (\frac{d_s\phi}{dt}) d^3p \} d^3q = \frac{dTS_{sources}}{dt} \qquad (105)$$

that, since T is not function of time (in stationary states) leads to

$$-\iiint_V \{ \int_{-\infty}^{+\infty}\int_{-\infty}^{+\infty}\int_{-\infty}^{+\infty} k \frac{(\phi-1)}{\phi} \rho^s (\frac{d_s\phi}{dt}) d^3p \} d^3q = \frac{dS_{sources}}{dt} \qquad (106)$$

if we displace the system from equilibrium in a stationary condition by imposing an external constraint, we have $\nabla\phi \neq 0$. Moreover, given that in a near-equilibrium state the variation of $\nabla\phi$ happens on a distance much larger than the local statistical system, hence, we can consider $\nabla\phi$ a locally constant field.

Making reference to the imposed external direction we can design $\nabla\phi_{//}$ as the component along this direction and $\nabla\phi_{\perp}$ the component perpendicular to it.

If we are in a stationary condition at local equilibrium, the fixed external constraint maintains $\nabla\phi_{//} \neq 0$ constant, while $\nabla\phi_{\perp}$ can fluctuate $\nabla\phi_{\perp} = 0 + \nabla\phi_{fluct}$, where $\nabla\phi_{fluct}$ is due to the statistical fluctuations.

Thence, being $Tr[D_q]$ definite positive, and being $\nabla\phi_{//}$ fix, from the SFED it follows that

$$|\frac{d_s\phi}{\delta t}| = D\nabla\phi \bullet \nabla\phi > 0 \qquad (107)$$

is minimum with respect the fluctuations of the system around $\nabla\phi_{\perp} = 0$. Thence, (106) represents the Prigogine's principle of minimum entropy production.

## 6.2 Spontaneous free energy increase in far from equilibrium steady-state transition

Without defining the characteristics of the systems and its boundary and initial conditions (that needs the specification of the local mean field of molecules, point by point) it is very difficult to gain insight about the general characteristics of far from equilibrium processes except for a very simple case.
To this end we consider below the electro-convective instabilities [9] carried out in quasi-isothermal condition.
In the case of quasi-isothermal system with no chemical reaction taking place we have already obtained that

$$\frac{dE_{res}}{dt} = \iiint_V \{ \int_{-\infty}^{+\infty} \int_{-\infty}^{+\infty} \int_{-\infty}^{+\infty} kT \frac{(\phi - 1)}{\phi} \rho^s (\frac{d_s \phi}{dt}) d^3 p \} d^3 q < 0 \quad (108)$$

where for electro-convective instabilities, has been assumed that each infinitesimal volume of fluid is at quasi-thermal equilibrium. On this base, $\Phi$ equals the free energy $F$, as well as $S^s$ the classical entropy $S$.
In the following we are going to show that in electroconvective instabilities the transitions to states with a higher amount of free energy are possible in far from equilibrium kinetics.
To this end, let's consider the overall system with the energetic reservoirs and the environment.
If the reservoir works reversibly so that

$$\frac{dE_{res}}{dt} = \frac{dF_{res}}{dt} \quad (109)$$

and the heat generated into the system is exchanged with the environment reversibly so that in a stationary condition it holds

$$\frac{dF_{sys}}{dt} = 0 \quad (110)$$

for the overall system it follows that

$$\frac{dF_{tot}}{dt} = \frac{dF_{res}}{dt} + \frac{dF_{sys}}{dt} = \iiint_V \{ \int_{-\infty}^{+\infty} \int_{-\infty}^{+\infty} \int_{-\infty}^{+\infty} kT \frac{(\phi - 1)}{\phi} \rho^s (\frac{d_s \phi}{dt}) d^3 p \} d^3 q < 0 \quad (111)$$

Actually, the condition $\frac{dE_{res}}{dt} = \frac{dF_{res}}{dt} < 0$ can be experimentally warranted in presence of transitions between metastable states. For instance in the in electro-convective instability this is obtained by posing a diode to the electric power in order to prevent that electric energy would flow back into the reservoir. This has been planned by Sawada and al. [9] but experiments showed that the current does not spontaneously change sign during transition between metastable states [22].
Let's consider the case that, far from equilibrium, the system makes a transition (at time t=0) from the metastable state (1) to another metastable one (2) so that for $t < 0^-$ and $t > 0^+$ we have the system in a stationary state for which it holds

$$\frac{dF_{sys\,1}}{dt} = \frac{dF_{sys\,2}}{dt} = 0 \quad (112)$$

so that by (111) we have ($t \neq 0$)

$$\begin{cases} \frac{dF_{tot}}{dt} < 0 \\ \frac{dF_{sys\,1}}{dt} = \frac{dF_{sys\,2}}{dt} = 0 \end{cases} \quad (113)$$

Given (111), in principle, the following cases are possible when the system makes the transition between states 1 and 2. For the forward transition $1 \rightarrow 2$ we have
case "a"

$$\begin{cases} \delta F_{tot\,(1\to 2)} < 0 \\ \delta F_{sys\,(1\to 2)} > 0 \end{cases} \tag{114}$$

case “b”

$$\begin{cases} \delta F_{tot\,(1\to 2)} < 0 \\ \delta F_{sys\,(1\to 2)} < 0 \end{cases} \tag{115}$$

For the backward transition $2 \to 1$ we analogously have
case “a”

$$\begin{cases} \delta F_{tot\,(2\to 1)} < 0 \\ \delta F_{sys\,(2\to 1)} < 0 \end{cases} \tag{116}$$

case “b”:

$$\begin{cases} \delta F_{tot\,(2\to 1)} < 0 \\ \delta F_{sys\,(2\to 1)} > 0 \end{cases} \tag{117}$$

When many metastable states exist with similar values of free energy dissipation, the system can switch between them due to external fluctuations (see section (7)). Therefore, given that transition between metastable can happen in both directions (if a fluctuation displaces the system from the first stationary state and is large enough then the maximum stochastic free energy dissipation brings the system to the new neighboring stationary state until another fluctuation lets the system to make the inverse pathway), it certainly follows that if in the direct transition happens that

$$\begin{cases} \delta F_{tot\,(1\to 2)} < 0 \\ \delta F_{sys\,(1\to 2)} < 0 \end{cases} \tag{118}$$

in the inverse one we have

$$\begin{cases} \delta F_{tot\,(2\to 1)} < 0 \\ \delta F_{sys\,(2\to 1)} > 0 \end{cases} \tag{119}$$

given that $\delta F_{sys\,(2\to 1)} = -\delta F_{sys\,(1\to 2)}$. Thence, considering the whole back and forth transition cycle, at least in one of the two directions, we have that

$$\delta F_{sys} > 0 \tag{120}$$

## 6.3 Entropy decrease in incompressible phases

Generally speaking, the increase of free energy does not also mean that the entropy of the system decreases. This depends by how much is the energy change in the transition.
Nevertheless due to the incompressibility of the fluid phase that makes null the volume energy variations, from the general point of view, in isothermal transitions in fluids (as the electro-convective ones) the free energy increase is almost due by the entropy decrease.
In fact given that

$$\delta F_{sys} \cong \delta E - \delta TS = \delta E_{int} + \delta E_{cin} - \delta TS \quad , \tag{121}$$

Since in isothermal transitions in fluids the temperature as well as the fluid density is practically constant, during the electro-convective transitions, for van der Waals type fluids where $E_{int} = E_{int}(T)$ it follows that $\delta E_{int} \cong 0$ and hence that

$$\delta F_{sys} \cong \delta E_{cin} - \delta TS > 0 \tag{122}$$

Moreover, given that for electro-convective instability in fluids [9] Sawada showed that at the transition between states with increasing numbers of vortexes (that gives a measure of the configurational order) the macroscopic kinetic energy of the molecules does not appreciably increases (or even decreases slightly with increasing order) [22] we have $\delta E_{cin} \cong 0$ and hence that

$$-\delta F_{sys} \cong T\delta S < 0 \qquad (123)$$

Therefore, the ability of a system to make back and forth transitions between metastable states as the consequence of the presence of fluctuations and the tendency to the maximum stochastic free energy dissipation, allows the spontaneous increase of order in far from equilibrium stationary states in incompressible fluid phase.

## 7. Discussion

By applying the QSHA model to the classical real gas phase it is possible to derive the far-from equilibrium kinetic equation in term of a Fokker–Plank equation.

The classical character of the gas phase comes by the fact that the mean inter-particle distance is larger than the quantum non-locality length that for L-J potential is of order of the molecular interaction distance $1{,}23\, r_0$ [16]. In fact, the L-J interacting gas phase has no inter-particle or long-range quantum correlations and it can be described by the Virial approximation considering only bimolecular Markovian collisions.

On the contrary, in solids and quantum fluids (e.g., super-fluids) the quantum potential range of interaction is bigger than the molecular interaction distance $r_0$ [16] so that the particles are quantum correlated.

For sufficiently rarefied real gas (van der Waals or mean field fluids included) the SQHA model allows for stationary state to transform the FPE (53 or 55) into a kinetic equation as a function of the phase space logarithm of the WFMV (LWFMV) $\phi$ that is well defined whether or not the system is at thermodynamic equilibrium.

If in the local thermodynamic equilibrium limit the LWFMV $\phi$ converges to the free energy of the system at least of a constant [19] (and for this reason called here *hydrodynamic free energy*) far from equilibrium the LWFMV allows to define a criterion of evolution.

If we instantaneously perturb the stationary state at time t = 0 (i.e., by changing the mean-field $\overline{H}$ and by generating a transient term $\delta \dot{x}_{trans} \neq 0$ in (77)) at t= 0$^+$ the kinetic equation (77) reads

$$\frac{d\phi}{dt} - < \dot{x}_{\overline{H}}' > \bullet \nabla\phi = (\delta \dot{x}_{trans} + < \dot{x}_s >) \bullet \nabla\phi = -D\nabla\phi \bullet \nabla\phi(1 + A + O(\nabla\phi^2)) \qquad (124)$$

For t>0, it follows that the system begins a transient during which $\nabla\phi$ and $(\delta \dot{x}_{trans} + < \dot{x}_s >)$ align themselves along each other (if a stationary state is achieved) ending with the $\nabla\phi'$ and $< \dot{x}_s ' >$ (ant-parallel to $\nabla\phi'$) so that the final kinetic equation reads

$$\frac{d\phi}{dt} - < \dot{x}_{\overline{H}}' > \bullet \nabla\phi' = < \dot{x}_s ' > \bullet \nabla\phi' = -D\nabla\phi' \bullet \nabla\phi'(1 + A + O(\nabla\phi'^2)) \qquad (125)$$

Depending by the boundary conditions and by the physical constants of the system, assigned a perturbation $\Delta\overline{H} = \overline{H'} - \overline{H}$ (or an external fluctuations) leading to a response $\delta \dot{x}_{trans}$, the final triplet of stationary values $\nabla\phi'$, $\nabla\phi'_{//}$ and $\nabla\phi'_{\perp}$ is assigned.

If the systems has no stability under the perturbation imposed, after the re-alignment $\nabla\phi'_{\perp}$ can differ from that one of the initial state, bringing the system to a different stationary configuration.

Given, for instance, a near-equilibrium initial stationary state of minimum entropy production (with $\nabla\phi_{\perp} = 0$) if the re-alignment of $\nabla\phi$ respect to $(\delta \dot{x}_{trans} + < \dot{x}_s >)$ generate a rotation of $\nabla\phi$ so that finally it results $\nabla\phi'_{\perp} \neq 0$, the new stationary state is not of minimum entropy production.

If the perturbation $\Delta\overline{H} = \overline{H'} - \overline{H}$ is small enough so that $\delta \overset{\bullet}{x}_{trans}$ is unable to generate a component perpendicular to $\nabla\phi_{//}$ and it is unable to rotate $\nabla\phi$, we end with $\nabla\phi'_{\perp} = 0$.

If $\delta \overset{\bullet}{x}_{trans}$ is given by an external imposed fix perturbation we may have $\nabla\phi'_{//} \neq \nabla\phi_{//}$. On the other hand if $\delta \overset{\bullet}{x}_{trans}$ is produced by an external fluctuations, that appears for an interval of time and then disappears, we may have $\nabla\phi'_{//} = \nabla\phi_{//}$.

If we apply a large perturbation $\Delta\overline{H} = \overline{H'} - \overline{H}$ so that we bring the system beyond the bifurcation point where many similar metastable state exist with different triplet of stationary values $\nabla\phi'$, $\nabla\phi'_{//}$, $\nabla\phi'_{\perp}$, when a large environmental fluctuation perturbs the system the new final triplets (after perturbation) may differ from the first one. This can happen because $\nabla\phi$ and $\delta \overset{\bullet}{x}_{trans}$ are coupled each other and $(\delta \overset{\bullet}{x}_{trans} + < \overset{\bullet}{x}_{s} >)$ follows the maximum SFED that practically is a maximum gradient criterion.

If the fluctuation generates a response $\delta \overset{\bullet}{x}_{trans}$ that leads to a new stochastic free energy gradient field $\nabla\phi'$ that is in the basin of convergence of a different stationary states the system will be driven toward it.

The objection that equation (66) is equivalent to introduce the local equilibrium condition is clearly not true. If the local equilibrium is set, then the WFMV gradient are small and (66) holds, but vice versa if the radiative coupling were null then the condition $\nabla\, ln[D^*] = 0$ would apply anyway whatever is the WFMV gradient. Therefore, in the case of weak coupling, the approximation (66) can extend itself beyond the local equilibrium condition.

The hydrodynamic free energy $\phi$ and the *hydrodynamic distribution function* $\rho^{s}$ are well defined in the far from equilibrium conditions and the evolution mechanism (62,66) leads to the maximal dissipation of the stochastic part of the hydrodynamic free energy stationary states.

This principle is not in contradiction with the preceding principle of Prigogine, Sawada and Malkus and Veronis, but agrees with them clarifying their controversial relationships.

The present model shows that in the case of a real gas or Marcovian fluids, with no chemical reactions at quasi-isothermal conditions, far from equilibrium the principle leads to maximum free energy dissipation given by Sawada and/or to the principle of maximum heat transfer given by Malkus and Veronis.

On the other hand, for the stationary state near-equilibrium before the sub-critical branch, the theory shows that the minimum SQHA-entropy production holds and that it disembogues into the Prigogine's principle. The minimum entropy production and the maximum statistical free energy dissipation are two different principia (two extremal criteria defined respect two different variations) but both descend by a unique coherent approach. The present treatment solves the apparently controversial aspect of these extremal principles.

It worth mentioning that the generation of order, via energy dissipation, is more efficient in fluids, because of their incompressibility, than in gasses leading the re-conciliation between physics and biology furnishing an explanation why the life was born in water.

Finally, the SQHA model suggests a scenario where the non-linearity plays a double key role: first to break the long distance quantum correlations, leading to the classically chaotic behavior, and then to elicit the appearance of order far from equilibrium in gas and Marcovian fluids. Under this light the characteristic of the universe and the life are deeply interconnected each other. Classical reality and life would have not been possible without the widespread non-linearity of physical interactions and the presence of vacuum fluctuations.

## 8. Conclusions

In the present work the SQHA is used as microscopic model to obtain the macro-scale non-equilibrium kinetics for a real gas of L-J interacting particles and its fluid phase.

In the case of particles whose mean distance is bigger than the quantum non-local length, that for L-J interaction potential is of order of the molecular interaction distance, the model shows that is possible to describe the SQHA evolution by means of a Fokker-Plank equation.

The novelty of the proposed proceeding consists in the introduction of the SQHA dynamics in deriving the Fokker-Plank kinetic equation. In gasses and Marcovian liquids, it is possible to define the kinetic equation of motion as a function of the phase space mean volume of the particles WFM whether the system is close or not to the local equilibrium.

For central symmetric potentials, the kinetic equation describing the evolution of the system maximizes the dissipation of the stochastic part of the SQHA-free energy.

In the case of a real gas with no chemical reactions and at quasi-isothermal conditions, the principle disembogues into the maximum free energy dissipation confirming the experimental outputs of electro-convective instability given by Sawada. In this case, the model shows that the transition to states with higher free energy can happen and that, in incompressible fluids, the increase of free energy is almost given by a decrease of entropy leading to the possibility of matter self-organization.

The model also shows that in the limit of local equilibrium the proposed principle disembogues into the Prigogine's one of minimum entropy production. The model is able to reconcile the minimum entropy production with the maximum free energy dissipation or the maximum heat transfer by Malkus and Veronis for atmosphere turbulence.

The theory shows that the maximum statistical free energy dissipation and the minimum entropy production are two independent principia not contradicting each other being defined respect to the variation of different variables.

## References

1. Prigogine, I. "Modération et transformations irreversibles des systemes ouverts". *Bulletin de la Classe des Sciences., Academie Royale de Belgique* **31**: (1945) 600–606.
2. Prigogine, I. *Étude thermodynamique des Phenomènes Irreversibles*, Desoer, Liege (1947).
3. Lavenda, B.H., *Thermodynamics of Irreversible Processes*, Macmillan, London, ISBN 0-333-21616-4 (1978).
4. Šilhavý, M., *The Mechanics and Thermodynamics of Continuous Media*, Springer, Berlin, ISBN 3-540-58378-5, (1997) p. 209.
5. Sawada, Y., A thermodynamic variational principle in nonlinear non-equilibrium phenomena, *Progress of Theoretical Physics 66:* (1981) *68-76.*
6. Malkus, W.V.R.; Veronis, G. "Finite amplitude cellular convection". *Journal of Fluid Mechanics* **4** (3): 225–260. Bibcode 1958JFM.....4..225M. doi:10.1017/S0022112058000410, (1958).
7. Onsager, L., "Reciprocal relations in irreversible processes, I". *Physical Review* **37** (4): 405–426. Bibcode 1931PhRv...37..405O. doi:10.1103/PhysRev.37.405 (1931).
8. Grandy, W.T., Jr, *Entropy and the Time Evolution of Macroscopic Systems*. Oxford University Press. ISBN 9780199546176 (2008)..
9. M. Suzuky and Y.Sawada, Relative stbilities of metastable states of convecting charged fluid systems by computer simulation, Phys. Rew. A, 27-1 (1983).
10. Gardiner, C.W., *Handbook of Stochastic Mechanics* (Springer-Verlag, Berlin and Heidelberg, 1985).
11. Peliti, L.: Path integral approach to birth-death processes on a lattice. J. Physique 46, 1469 (1985);
12. Madelung, E.: Quanten theorie in hydrodynamische form (Quantum theory in the hydrodynamic form). Z. Phys. 1926; 40, 322-6, German.
13. I. Bialynicki-Birula, M. Cieplak and J.Kaminski, *Theory of Quanta*, Oxford University Press, Ny (1992).
14. Weiner, J.H., *Statistical Mechanics of Elasticity* (John Wiley & Sons, New York, 1983), p. 316-317.
15. Chiarelli, P., "Can fluctuating quantum states acquire the classical behavior on large scale?" arXiv: 1107.4198 [quantum-phys] (2012), submitted for pblication on Foud. of Phys.

16. Chiarelli, P., "Quantum to classical phases transition in the stochastic hydrodynamic analogy: a possible connection between the maximum of density at He lambda point and that one at water-ice phase transition", in press on Phys. Rew. & Res. Int., (2013).
17. Gardiner, C.W., *Handbook of Stochastic Mechanics* (Springer-Verlag, Berlin and Heidelberg, 1985) pp. 301-314.
18. Rumer, Y. B., Ryvkin, M. S., *Thermodynamics, Statistical Physics, and Kinetics* (Mir Publishers, Moscow, 1980), p. 325-334.
19. Chiarelli, P., "Far from equilibrium maximal principle leading to matter self-organization" arXiv: 1303.1772 [cond-mat.stat-mech ] (2013).
20. Ibid 18 pp. 465-474.
21. Ibid 18 p. 496.
22. Ibid 22, pp. 482,486; fig 5 and fig. 9.

# Appendix A

### From dynamics to statistics

### The classic-stochastic and the quantum-deterministic aspects of the SQHA probability distribution

The existence of the quantum non-locality length $\lambda_L$ and hence of a scale-transition between the quantum and the classic dynamics confers to the PTF of ρ (namely $P(\rho)$) both the quantum and the statistical character.
Once the $P(\rho)$ of the SPDE (13) is defined, both the quantum wave equation on "microscopic" scale and the statistical distribution on huge-scale are defined.
When the quantum coherence length $\lambda_c$ goes to infinity (with respect the scale of our system or description) $P(\rho)$ tends to the Dirac function $\delta(\rho - \rho_{(deterministic)})$ so that ρ tends to $\rho_{(deterministic)}$ and the SQHA converges to the quantum mechanics. In this case, the PDF ρ has the full quantum meaning given by (4-5) and actually is a "quantum dynamical " distribution.
On the other hand, when $\lambda_L$ is very small compared to the physical length $\Delta L$ of the problem (e.g., mean particle distance or free molecular path), the classical stochastic dynamics (24-29) arises.
When we deal with a system of a huge number of (non-linearly interacting) particles with a finite interaction distance [41] (i.e, $r_0$ for L-J potentials), each coarse-graining cell with a side $\Delta\Omega_q{}^s >> \Delta L >> \lambda_L, r_0$ (containing a large number of molecules) can constitute a local system. This because in the thermodynamical limit (infinite system volume) the quantum correlations involve a small fraction of molecules in a thin layer at the $\Delta\Omega_q{}^s$-boundary.
Therefore, when superficial effects can be disregarded with respect to those of the bulk, the overall system can be ideally subdivided into a large number of quantum uncorrelated randomly distributed $\Delta\Omega_q{}^s$-subsystems.
In this case it is possible to write the statistical distribution of those $\Delta\Omega_q{}^s$-copies in terms of operators applied to the "mother distribution" $P(\rho)$. This is warranted by the fact that once the evolution of the SQHA probability $P(\rho)$ is defined, it also defines the evolution of the corrispective (statistical) distribution on large-scale.
It is also worth noting that the $P(\rho)$-derived statistical distribution remains defined regardless the establishment of the local thermodynamic equilibrium.

### The coarse-grained SQHA statistical distribution

Here, we derive the statistical distribution from the SHQA dynamics distribution by subdividing the system in cells of side $\Delta\Omega_q{}^s$.

In order to have independent random $\Delta\Omega_q{}^s$-copies of the system when the (local) equilibrium is set, in addition to the conditions: (1) $\Delta\Omega_q{}^s >> \Delta L$ (where $\Delta L$ is the physical length on molecular scale (e.g., the mean molecular distance)) and (2) $\lambda_L << \Delta\Omega_q{}^s$, we need that the molecules in $\Delta\Omega_q{}^s$ interact with the particles out of it, just through the border of $\Delta\Omega_q{}^s$ for a layer whose characteristic length is much smaller than $\Delta\Omega_q{}^s$ radius.

Assuming both $\lambda_L << \Delta\Omega_q{}^s$ and $\Delta L << \Delta\Omega_q{}^s$ (given that $r_0 < \Delta L$ or even $r_0 << \Delta L$ as in the gas phase), the particles on $\Delta\Omega_q{}^s$ border are an infinitesimal fraction of those ones contained in its volume. Hence, in the limit of infinite (sufficiently large) $\Delta\Omega_q{}^s$ volume, the superficial effects (quantum ones included) can be disregarded so that the $\Delta\Omega_q{}^s$ domains tend to be de-coupled and quantum uncorrelated each others.

Under these conditions, since the $\Delta\Omega_q{}^s$-systems are constituted by a sufficiently large number of non-linearly interacting particles and, hence, classically chaotic, they can be assumed to evolve with random initial conditions because the correlation with their initial state decays quickly.

Therefore, when the local equilibrium is set, the $\Delta\Omega_q{}^s$-domains are random copies of each other and build up the grand canonical ensemble .

For sake of completeness, it must be said that the prescriptions " (2)"it is of fundamental importance in setting the classical description (i.e., $\lambda_L < \infty$) as well as the domains $\Delta\Omega_q{}^s$ of finite side.

On the contrary, in the case of perfectly harmonic solids, the quantum potential range of interaction $\lambda_L$ will result infinite and the local means cannot be defined. Since we require the $\Delta\Omega_q{}^s$-cell length much larger than $\lambda_L$, in this case it would comprehend the entire system and we cannot speak in term of local stochastic means but only in term of the quantum means of the entire system.

Actually, for a real solid the intermolecular potential is of L-J type. In this case, the harmonic interaction extends itself to the nearest molecules but not to infinity and $\lambda_L$ is in the reality finite [see [p rew& res 2] N].

We can warrant the above prescriptions by restraining ourselves to the sufficiently general case, to be of interest, of finite-range potentials that have a rapid decreasing Mayers functions [18] as for gas and van der Waals fluids or, more generally for L-J potentials for which we have

$\lambda_L \cong 0{,}23 r_0 < \Delta L << \Delta\Omega^s$ [16].

We follow here the same approach for interacting particles given in ref. [18].

If $\Delta N_k$ is the number of the groups of k-molecules in $\Delta\Omega_q{}^s$ and $\Delta n_{kjh}$ the number of the groups of k-molecules in the domains $\Delta\Omega_{jh} = ((j-1)\Delta q < q < j\Delta q,\ (h-1)\Delta p < p < h\Delta p)$ contained in $\Delta\Omega_q{}^s$, we obtain

$$\Delta N_k = \sum_{j\in\Delta\Omega_q{}^s} \sum_{h=-\infty}^{+\infty} \Delta n_{kjh} \qquad \text{(A1)}$$

where

$$\Delta n_{kjh} = \sum_i \int_{(j-1)\Delta q_1}^{j\Delta q_1} .... \int_{(j-1)\Delta q_{3k}}^{j\Delta q_{3k}} \left( \int_{(h-1)\Delta p_1}^{h\Delta p_1} .... \int_{(h-1)\Delta p_{3k}}^{h\Delta p_{3k}} P_{i(k)}(\rho) d^{3k}p \right) d^{3k}q = \sum_i n_{ikjh} \qquad \text{(A2)}$$

where V is the system volume, *i(k)* is the index of the i-th group of k-molecules in the entire system and $\boldsymbol{P}_{i(k)}$ is the projector operator for the *i(k)*-th group that reads

$$P_{i(k)} = \int_{-\infty}^{+\infty} ....\int_{-\infty}^{+\infty} d^{3k}q_1 d^{3k}q_{j\neq i} d^{3k}q_{3\frac{n!}{(n-k)!k!}} \int_{-\infty}^{+\infty} ....\int_{-\infty}^{+\infty} d^{3k}p_1 d^{3k}p_{j\neq i} d^{3k}p_{3\frac{n!}{(n-k)!k!}}$$

(A3)

It must be said that being ρ the quantum probability (18), it implicitly accounts for the indistinguishability of particles.

The density of states of groups of k molecules on phase space domain $\Delta\Omega_{jh}$ reads

$$\rho^s{}_{k(q_j,p_h)} = \frac{\Delta n_{khj}}{\Delta q^{3k} \Delta p^{3k}}$$

$$= \Delta q^{-3k} \Delta p^{-3k} \sum_i \int_{(j-1)\Delta q_1}^{j\Delta q_1} .... \int_{(j-1)\Delta q_{3k}}^{j\Delta q_{3k}} \left( \int_{(h-1)\Delta p_1}^{h\Delta p_1} .... \int_{(h-1)\Delta p_{3k}}^{h\Delta p_{3k}} P_{i(k)}(\rho) d^{3k}p \right) d^{3k}q \qquad \text{(A6)}$$

It is worth noting that the distribution (A7) si not generally a statistical distribution. It acquires the statistical character when the $\Delta\Omega_q{}^s$ -systems is constituted by a sufficiently large number of sub-systems $\Delta\Omega_q{}^s{}_j = \sum_{h=-\infty}^{h=+\infty} \Delta\Omega_{jh}$ not–correlated eachother evolving with random initial conditions (this can happen in a system of non-linear classically chaotic particles where $\lambda_c$ *and* $\lambda_L << \Delta q << \Delta\Omega_q{}^s$ ).

Given the energy function for a group of k molecules

$$E_k = \sum_i E_{i(k)} = \sum_i \frac{p_{n(i(k))} p_{n(i(k))}}{2m} + U_{k(i)} \qquad \text{(A10)}$$

where $p_{n(i(k))}$ is the momentum of the n-th molecule of the i-th group of k molecules and $U_{k(i)}$ is the potential energy of the i-th group of k molecules. Therefore, the mean value in $\Delta\Omega_{jh}$ follows

$$< E_k >= \frac{\sum_i n_{ikhj} E_{i(k)}}{\sum_i n_{ikhj}} = \frac{\sum_i \frac{n_{ikhj}}{\Delta q^{3k} \Delta p^{3k}} E_{i(k)} \Delta q^{3k} \Delta p^{3k}}{\sum_i n_{ikhj}}$$

$$= \frac{\int \underset{\Delta\Omega}{....} \int \rho^s{}_k E_{i(k)} d^{3k} q \, d^{3k} p}{\Delta n_{khj}}$$

(A11)

where

$$< E_k >_{(q,p)} = \frac{\sum_i \int_{(j-1)\Delta q_1}^{j\Delta q_1} \int_{(j-1)\Delta q_{3k}}^{j\Delta q_{3k}} \int_{(h-1)\Delta p_1}^{h\Delta p_1} \int_{(h-1)\Delta p_{3k}}^{h\Delta p_{3k}} P_{i(k)}(\rho) E_{i(k)} d^{3k} p d^{3k} q}{\sum_i \int_{(j-1)\Delta q_1}^{j\Delta q_1} \int_{(j-1)\Delta q_{3k}}^{j\Delta q_{3k}} \int_{(h-1)\Delta p_1}^{h\Delta p_1} \int_{(h-1)\Delta p_{3k}}^{h\Delta p_{3k}} P_{i(k)}(\rho) d^{3k} p d^{3k} q}$$

$$< E >_{(q,p)} = \frac{\sum_k \Delta n_{khj} < E_k >}{\sum_k \Delta n_{khj}} \qquad \text{(A12)}$$

Moreover, defining the operator $\mathrm{O}^s{}_{p_{i(k)}}$ such as

$$\mathrm{O}^s{}_{p_{i(k)}} = \Delta q^{-3k} \Delta p^{-3k} \sum_i \int_{(j-1)\Delta q_1}^{j\Delta q_1} \int_{(j-1)\Delta q_{3k}}^{j\Delta q_{3k}} \int_{(h-1)\Delta p_1}^{h\Delta p_1} \int_{(h-1)\Delta p_{3k}}^{h\Delta p_{3k}} P_{i(k)} d^{3k} p d^{3k} q \qquad \text{(A16)}$$

we can formally link the coarse-grained quantities to the "mother probability distribution" ρ in a synthetic manner as follows

$$\mathrm{O}^s{}_{p_{i(k)(\rho)}} = \rho^s{}_{k(q,p)} \qquad \text{(A17)}$$

$$\mathrm{O}^s{}_{p_{i(k)(\rho)}} (E_k) = \rho^s{}_{k(q,p)} < E_k >$$

(A18)

and, finally,

$$< E >_{(q,p)} = \frac{\sum_k \Delta n_{khj} < E_k >}{\sum_k \Delta n_{khj}} = \frac{\sum_k \Omega^s{}_{p_{i(k)(\rho)}} (E_k)}{\sum_k \Omega^s{}_{p_{i(k)(\rho)}}} \qquad \text{(A19)}$$

The summation over all the configurations ρ with probability $P(\rho, \rho')$ leads to the re-defined quantity

$$\rho^s{}_{k(q,p)} = \int \mathrm{O}^s{}_{p_{i(k)(\rho)}} P(\rho,\rho')d\rho' \qquad \text{(A20)}$$

$$< E >_{(q,p)} = \frac{\int \sum_k \mathrm{O}^s{}_{p_{i(k)(\rho)}} (E_k) P(\rho,\rho')d\rho'}{\int \sum_k \mathrm{O}^s{}_{p_{i(k)(\rho)}} P(\rho,\rho')d\rho'} \qquad \text{(A21)}$$

In many macroscopic problem we can approximate the PTF $P(\rho)$ as a δ-peaked function such as $P(\rho) \cong \delta(\rho' - \rho)$ so that (A20) leads to (A17).

### The local equilibrium limit

As far as it concerns the above distribution (A20) as well as the mean energy (A21), it is worth noticing that no hypothesis on thermodynamical equilibrium has been introduced to obtain them so far.
If the characteristic length over which the thermodynamical gradients generate appreciable variations is much bigger than the system dimension, so that the equilibrium can be assumed, the $\Delta\Omega^s$-domains in absence of external fields (otherwise appropriate thermodynamical potentials can be defined) become (random) copies of each others giving rise to the canonical ensemble. In force of this "equalization process" (when it happens), the coarse-grained quantity (A20-A21) can refer to those of the canonical ensemble of the $\Delta\Omega^s$ –random–copies, converging to the classical expressions of the statistics of equilibrium. By utilizing the definition of the partition function $Z_k$

$$Z_k = \frac{1}{V} \int_{-\infty}^{+\infty} .... \int_{-\infty}^{+\infty} \rho^{eq}{}_k d^{3k}q \, d^{3k}p$$

we obtain that

$$lim_{TE} \int_{-\infty}^{+\infty} .... \int_{-\infty}^{+\infty} \rho^s{}_k d^{3k}q \, d^{3k}p = N_k = Z_k{}^{-1} \frac{N_k}{V} \int_{-\infty}^{+\infty} .... \int_{-\infty}^{+\infty} \rho^{eq}{}_k d^{3k}q \, d^{3k}p$$

where $N_k = \sum_k \Delta N_k$ and hence that

$$lim_{TE} \, \rho^s{}_k = Z_k{}^{-1} \frac{N_k}{V} \rho^{eq}{}_k \qquad \text{(A22)}$$

It must be noted that in order to obtain the familiar distributions on the basis of the molecular approach, the physical length $\Delta L$ has to be set equal to the typical molecular dimension.

### Idel gas

For ideal gas (i.e., punctual particles with $r_0 \rightarrow 0$) the only relevant distribution is for k = 1 so that (??) redas

$$lim_{TE}\ \rho^s{}_1 = Z^{-1}\frac{N}{V}\rho^{eq}, \tag{A22}$$

for a real gas the only relevant values of k are k = 1, 2, while for condensed phases k > 1 (practically, for L-J potentials with strong repulsive core, it can be taken values of k about those of the coordination number of the elemental cell).

Given that the interaction distance for the Hamiltonian L-J potential is of order of $r_0$ as well as for the quantum pseudo-potential (of order of $r_0 + \lambda_l$) [phys rev & res int], in a sufficiently rarefied gas phase, particles can be assumed independent (the ideal gas approximation assumes $r_0$ =0) and the SQHA-WFM distribution can be factorized as

$$\rho = \prod_i \rho_i$$

and the one-particle group projector $P_{i(1)}$ reads

$$P_{i(1)} = \int_{-\infty}^{+\infty}....\int_{-\infty}^{+\infty} d^3q_1 d^3q_{j\neq i} d^3q_{3n} \int_{-\infty}^{+\infty}....\int_{-\infty}^{+\infty} d^3p_1 d^3p_{j\neq i} d^3p_{3n}$$

leading to

$$P_{i(1)}(\rho) = \rho_i$$

and

$$\rho^s(q_j, p_h) = \frac{\sum_i \int_{(j-1)\Delta q_1}^{j\Delta q_1} .... \int_{(j-1)\Delta q_3}^{j\Delta q_3} \left( \int_{(h-1)\Delta p_1}^{h\Delta p_1} .... \int_{(h-1)\Delta p_3}^{h\Delta p_{3k}} \rho_i d^3p \right) d^3q}{\Delta q^3 \Delta p^3} = \frac{\sum_i \int ...\int_{\Delta\Omega} \rho_i d^3p d^3q}{\Delta\Omega}$$

$$\mathrm{O}^s{}_{P_{i(1)(\rho)}} = \rho^s(q,p) \tag{A17}$$

$$\mathrm{O}^s{}_{P_{i(1)(\rho)}}(E) = \rho^s(q,p) < E >$$

Using the definition of the projector operator $P_{i(1)}$, the number of particles $\Delta n_{1jh}$ in $\Delta\Omega_{jh}$ can be expressed as (A1)

$$\Delta n_{1jh} = \sum_i \int_{(j-1)\Delta q_1}^{j\Delta q_1} \int_{(j-1)\Delta q_{3k}}^{j\Delta q_{3k}} \int_{(h-1)\Delta p_1}^{h\Delta p_1} \int_{(h-1)\Delta p_{3k}}^{h\Delta p_{3k}} \rho_i d^3p d^3q$$

with the normalization condition $\iiint d^3q \iiint d^3 p\rho_i = 1 \; \forall$ i.
The summation over all the configurations ρ with probability $P(\rho, \rho')$ leads to the re-defined quantities

$$\rho^s(q,p) = \int \mathrm{O}^s p_{i(1)(\rho)} P(\rho, \rho')d\rho' \tag{A20}$$

$$< E >_{(q,p)} = \frac{\int \mathrm{O}^s p_{i(1)(\rho)} (E_k) P(\rho, \rho')d\rho'}{\int \mathrm{O}^s p_{i(1)(\rho)} P(\rho, \rho')d\rho'} \tag{A21}$$

# APPENDIX B

## Thermodynamic equilibrium

Let' find now the equilibrium quantities $\rho^{eq}$ and $\lim_{TE} D = D_{TE}$ .

Being the thermodynamic equilibrium the stationary state with null dissipation (i.e., $\frac{\partial \rho^s}{\partial t} = 0$, $\frac{d\rho^s}{dt} = 0$ and $\nabla\phi = 0$ ) by (78-79) it follows that

$$\lim_{TE} < \overset{\bullet}{x}_s > = -D\nabla\phi(1 + A + O(\nabla\phi^2)) = 0 \tag{80}$$

and, being $D > 0$ , therefore, in absence of external field (for isotropic condition) that

$$\lim_{TE} \phi = \phi_{TE} = Constant$$

where $\lim_{TE}$ indicates the establishing of local thermodynamic equilibrium.
Moreover, given from (??) that

$$\rho^s = h^{-3}\frac{D^*}{D} exp[\phi]$$

it follows that

$$D = h^{-3} D^* exp[\phi + \frac{S^s}{k}]$$

and hence

$$\lim_{TE} D = D_{TE} = h^{-3} D^* exp[\phi_{TE} + \frac{S^{eq}}{k}]$$

where

$$\lim_{TE} S^s = S^{eq}$$

$$\phi_{TE} = ln[h^{-3}D^*] + ln[D_{TE}] - \frac{S^{eq}}{k}$$

that by posing

$$ln[h^{-3}D^*] = \phi_0$$

reads

$$\phi_{TE} - \phi_0 = ln[D_{TE}] - \frac{S^{eq}}{k} \qquad \text{(*eq)}$$

leading to

$$D_{TE} = exp[\phi_{TE} - \phi_0 + \frac{S^{eq}}{k}]$$

Finally, as well known [], to derive the Maxwell-Boltzmann equilibrium distribution from the Fokker – Plank one additional information has necessarily to be introduced (i.e., the linear empirical relations between gradients and fluxes (i.e., null fluxes for null gradients).
In the present approach, we impose the "equalization condition " on the phase space WFM volume $\phi$ assuming that there is translation invariance of $\phi$ and zero fluxes of it .
Given the thermodynamical equilibrium as the stationary state with null dissipation (constant free energy) and null net fluxes of free energy at the boundary (i.e., $\frac{d\Phi_{sup}}{dt} = 0$ ), by (?) in appendix C it follows that

$$\lim_{TE} k\frac{dT\phi}{dt} - \frac{d(E_{cin} + E_{int})}{dt} + \frac{dTS^s{}_{vol}}{dt} = -\iiint_V \int_{-\infty}^{+\infty}\int_{-\infty}^{+\infty}\int_{-\infty}^{+\infty} kT\frac{(\phi-1)}{\phi}\rho^s(\frac{d_s\phi}{dt})d^3p\ d^3q = 0$$

and that

$$\lim_{TE} k\frac{dT\phi}{dt} = \lim_{TE}(\frac{dE}{dt} - \frac{dTS^s{}_{vol}}{dt}). \qquad \text{(?'?)}$$

That , fnally integrating (?'?) ed by using (*eq), leads to

$$\phi_{TE} - const = \frac{E_{TE}}{kT} - \frac{S^s{}_{vol\,TE}}{k} = ln[D_{TE}] - \frac{S^{eq}}{k}$$

and, hence, that

$$D_{TE} = exp[\frac{E_{TE}}{kT}]$$

and that

$$\phi_{TE} - \phi_0 = \frac{E_{TE}}{kT} - \frac{S^{eq}}{k}$$

represents the themodynamical free energy. Moreover, by using (69) the above relations ends with

$$\lim_{TE} \lim_{\Delta L >> \lambda_c, \lambda_L} \rho^s = \rho^{eq} = \lim_{TE} h^{-3} \frac{D^*}{D} exp[\phi]$$

$$= \lim_{TE} exp[\phi - \phi_0] \, exp[-\frac{E}{kT}] = exp[\phi_{TE} - \phi_0] \, exp[-\frac{E_{TE}}{kT}]$$

that is the Maxwell-Boltzmann equilibrium distribution.

Finally, it is interesting to see that the variation of the proportionality constant α in (?) brings to the change of the equilibrium free energy by a constant. In fact, given that

$$\rho^{eq} = h^{-3} \alpha \, exp[\phi_{TE}] = h^{-3} \alpha' \, exp[\phi'_{TE}]$$

it follows that

$$ln[\frac{\alpha}{\alpha'}] = \phi' - \phi$$

### Out of local thermodynamic equilibrium

Out of equilibrium we can set

$$\rho^s = exp[\phi - \phi_0] \, exp[-\frac{E}{kT}] = exp[\phi - \phi_{TE}] \, exp[-\frac{E - E_{TE}}{kT}] \, exp[\phi_{TE} - \phi_0] \, exp[-\frac{E_{TE}}{kT}]$$

$$= exp[\Delta\phi] \, exp[-\frac{\Delta E}{kT}] \, exp[\phi_{TE} - \phi_0] \, exp[-\frac{E_{TE}}{kT}] = \rho^{eq} \, exp[-\frac{\Delta S}{k}]$$

with

$$\Delta\phi = \phi - \phi_{TE}$$

$$\Delta E = E - E_{TE}$$

$$\frac{\Delta S}{k} = -\Delta\phi + \frac{\Delta E}{kT}$$

$$\Delta S = -k \, ln[\frac{\rho^s}{\rho^{eq}}]$$

$$\phi = \phi_{TE} + \frac{\Delta E}{kT} - \frac{\Delta S}{k}$$

The local mean field energy $E$ can be ideally seen as composed by two parts, one given by the (local) equilibrium one $E_{TE}$ and the other one $\Delta E$ given by the deviation from it.

## APPENDIX C

### Spatial kinetic equations

we transform the local law (I3.1) into a spatial one over a finite volume V.

$$\underline{\Upsilon} = \frac{\int_{-\infty}^{+\infty}\int_{-\infty}^{+\infty}\int_{-\infty}^{+\infty}\rho^s \Upsilon d^3 p}{\int_{-\infty}^{+\infty}\int_{-\infty}^{+\infty}\int_{-\infty}^{+\infty}\rho^s d^3 p} \tag{104}$$

its sptatial density:

$$n\underline{\Upsilon} = \int_{-\infty}^{+\infty}\int_{-\infty}^{+\infty}\int_{-\infty}^{+\infty}\rho^s \Upsilon d^3 p \tag{105}$$

and first moment

$$n\underline{\Upsilon \dot{q}} = \int_{-\infty}^{+\infty}\int_{-\infty}^{+\infty}\int_{-\infty}^{+\infty}\rho^s \Upsilon < \dot{q} > d^3 p \tag{106}$$

by using the motion equation (75) it is possible to obtain the spatial differential equation:

$$\begin{aligned} &\partial_t n\underline{\Upsilon} + \nabla \bullet n\underline{\Upsilon \dot{q}} - \int_{-\infty}^{+\infty}\int_{-\infty}^{+\infty}\int_{-\infty}^{+\infty} \rho^s \{ \partial_t \Upsilon + < \dot{x}_{<H>} > \bullet \nabla \Upsilon \} d^3 p \\ &= \int_{-\infty}^{+\infty}\int_{-\infty}^{+\infty}\int_{-\infty}^{+\infty} \Upsilon \{ \nabla \bullet \rho^s D \nabla \phi (1 + A + O(\nabla \phi^2 ))\} d^3 p \end{aligned} \tag{107}$$

That by choosing

$$\Upsilon = kT\phi \tag{108}$$

where T is the "mechanical" temperature defined as

$$T = \gamma \frac{< E_{cin} + E_{pot} >}{k} = \gamma ( \frac{\frac{< p_i >< p_i >}{2m} + < V_{i \neq j} >}{k} )$$

where $\gamma$ depends by the system under consideration and can be calculated at thermodynamical equilibrium (e.g., 2/3 for ideal gas, 1/3 for classical oscillators etc.). Moreover, by posing

$$\Upsilon = \Upsilon_{TE} + \Delta \Upsilon = E - TS_{TE} + \Delta \Upsilon$$

it follows that

$$\partial_t n\underline{Y} + \nabla \bullet n\underline{Y\,\dot{q}} - \int_{-\infty}^{+\infty}\int_{-\infty}^{+\infty}\int_{-\infty}^{+\infty} \rho^s \{\partial_t Y_{TE} + <\dot{x}_{<H>}> \bullet \nabla Y_{TE}\} d^3p$$

$$- \int_{-\infty}^{+\infty}\int_{-\infty}^{+\infty}\int_{-\infty}^{+\infty} \rho^s \{\partial_t \Delta Y + <\dot{x}_{<H>}> \bullet \nabla(\Delta Y)\} d^3p$$

$$= \int_{-\infty}^{+\infty}\int_{-\infty}^{+\infty}\int_{-\infty}^{+\infty} Y\{\nabla \bullet \rho^s D\nabla\phi(1 + A + O(\nabla\phi^2))\} d^3p$$

that

$$\partial_t n\underline{Y} + \nabla \bullet n\underline{Y\,\dot{q}} - \int_{-\infty}^{+\infty}\int_{-\infty}^{+\infty}\int_{-\infty}^{+\infty} \rho^s \{\partial_t E + <\dot{x}_{<H>}> \bullet \nabla E\} d^3p$$

$$- \int_{-\infty}^{+\infty}\int_{-\infty}^{+\infty}\int_{-\infty}^{+\infty} \rho^s \{\partial_t \Delta Y + <\dot{x}_{<H>}> \bullet \nabla(\Delta Y)\} d^3p$$

$$= - \int_{-\infty}^{+\infty}\int_{-\infty}^{+\infty}\int_{-\infty}^{+\infty} \rho^s \{\partial_t TS + <\dot{x}_{<H>}> \bullet \nabla TS\} d^3p + \int_{-\infty}^{+\infty}\int_{-\infty}^{+\infty}\int_{-\infty}^{+\infty} Y\{\nabla \bullet \rho^s D\nabla\phi(1 + A + O(\nabla\phi^2))\} d^3p$$

that

$$\partial_t n\underline{Y} + \nabla \bullet n\underline{Y\,\dot{q}} - \int_{-\infty}^{+\infty}\int_{-\infty}^{+\infty}\int_{-\infty}^{+\infty} \rho^s \{\frac{dE}{dt}\} d^3p - \int_{-\infty}^{+\infty}\int_{-\infty}^{+\infty}\int_{-\infty}^{+\infty} \rho^s \{\partial_t \Delta Y + <\dot{x}_{<H>}> \bullet \nabla(\Delta Y)\} d^3p$$

$$= - \int_{-\infty}^{+\infty}\int_{-\infty}^{+\infty}\int_{-\infty}^{+\infty} \rho^s \{\partial_t TS + <\dot{x}_{<H>}> \bullet \nabla TS\} d^3p + \int_{-\infty}^{+\infty}\int_{-\infty}^{+\infty}\int_{-\infty}^{+\infty} Y\{\nabla \bullet \rho^s D\nabla\phi(1 + A + O(\nabla\phi^2))\} d^3p$$

that

$$\partial_t n\underline{Y} + \nabla \bullet n\underline{Y\,\dot{q}} - \int_{-\infty}^{+\infty}\int_{-\infty}^{+\infty}\int_{-\infty}^{+\infty} \rho^s \{<\dot{p}> \bullet <\dot{q}>\} d^3p - \int_{-\infty}^{+\infty}\int_{-\infty}^{+\infty}\int_{-\infty}^{+\infty} \rho^s \{\partial_t \Delta Y + <\dot{x}_{<H>}> \bullet \nabla(\Delta Y)\} d^3p$$

$$= - \int_{-\infty}^{+\infty}\int_{-\infty}^{+\infty}\int_{-\infty}^{+\infty} \rho^s \{\partial_t TS + <\dot{x}_{<H>}> \bullet \nabla TS\} d^3p + \int_{-\infty}^{+\infty}\int_{-\infty}^{+\infty}\int_{-\infty}^{+\infty} Y\{\nabla \bullet \rho^s D\nabla\phi(1 + A + O(\nabla\phi^2))\} d^3p$$

that

$$\partial_t n\underline{Y} + \nabla \bullet n\underline{Y\dot{q}} - F \bullet \underline{\dot{q}} - \int_{-\infty}^{+\infty}\int_{-\infty}^{+\infty}\int_{-\infty}^{+\infty} \rho^s \{ \partial_t \Delta Y + < \dot{x}_{<H>} > \bullet \nabla(\Delta Y) \} d^3 p$$

$$= -\int_{-\infty}^{+\infty}\int_{-\infty}^{+\infty}\int_{-\infty}^{+\infty} \rho^s \{ \partial_t TS + < \dot{x}_{<H>} > \bullet \nabla TS \} d^3 p + \int_{-\infty}^{+\infty}\int_{-\infty}^{+\infty}\int_{-\infty}^{+\infty} Y\{ \nabla \bullet \rho^s D \nabla \phi (1 + A + O(\nabla \phi^2)) \} d^3 p$$

$$\partial_t n\underline{Y} + \nabla \bullet n\underline{Y\dot{q}} - F \bullet \underline{\dot{q}} - \int_{-\infty}^{+\infty}\int_{-\infty}^{+\infty}\int_{-\infty}^{+\infty} \rho^s \{ \partial_t \Delta Y + < \dot{x}_{<H>} > \bullet \nabla(\Delta Y) \} d^3 p$$

$$= -\int_{-\infty}^{+\infty}\int_{-\infty}^{+\infty}\int_{-\infty}^{+\infty} \rho^s \{ T\partial_t S + < \dot{x}_{<H>} > \bullet T\nabla S \} d^3 p - \int_{-\infty}^{+\infty}\int_{-\infty}^{+\infty}\int_{-\infty}^{+\infty} \rho^s \{ S\partial_t T + < \dot{x}_{<H>} > \bullet S\nabla T \} d^3 p$$

$$+ \int_{-\infty}^{+\infty}\int_{-\infty}^{+\infty}\int_{-\infty}^{+\infty} Y\{ \nabla \bullet \rho^s D \nabla \phi (1 + A + O(\nabla \phi^2)) \} d^3 p$$

Where the force on unit volume $F$ reads $F = < \dot{p} >_{(q)}$ and for Hamiltonian potentials that are not function of momenta can be brought out of the integral, and where $\underline{\dot{q}} = \int_{-\infty}^{+\infty}\int_{-\infty}^{+\infty}\int_{-\infty}^{+\infty} \rho^s < \dot{q} > d^3 p$.

Moreover, given that for elastic molecular collisions (e.g., no chemical reactions) H is conserved so that

$$\nabla \bullet < \dot{x}_{<H>} > = \frac{\partial^2 H}{\partial p_i \partial q_i} - \frac{\partial^2 H}{\partial q_i \partial p_i} = 0,$$

with the help of (73,77) it follows that

$$\partial_t \rho^s + \nabla \bullet \rho^s ( < \dot{x}_{<H>} > + < \dot{x}_s > ) = 0$$

$$\partial_t \rho^s + < \dot{x}_{<H>} > \nabla \rho^s = -\nabla \bullet ( \rho^s < \dot{x}_s > )$$

$$\partial_t \rho^s + < \dot{x}_{<H>} > \nabla \rho^s = \nabla \bullet ( \rho^s D \nabla \phi (1 + A + O(\nabla \phi^2)))$$

we obtain

$$\int_{-\infty}^{+\infty}\int_{-\infty}^{+\infty}\int_{-\infty}^{+\infty} \rho^s \{T\partial_t S + <\overset{\bullet}{x}_{<H>}> \bullet T\nabla S_{TE}\} d^3 p$$

$$= \int_{-\infty}^{+\infty}\int_{-\infty}^{+\infty}\int_{-\infty}^{+\infty} \rho^s \{T\partial_t S^s + <\overset{\bullet}{x}_{<H>}> \bullet T\nabla S^s\} d^3 p - \int_{-\infty}^{+\infty}\int_{-\infty}^{+\infty}\int_{-\infty}^{+\infty} \rho^s \{T\partial_t \Delta S + <\overset{\bullet}{x}_{<H>}> \bullet T\nabla \Delta S\} d^3 p$$

$$= \int_{-\infty}^{+\infty}\int_{-\infty}^{+\infty}\int_{-\infty}^{+\infty} kT\{\partial_t \rho^s + <\overset{\bullet}{x}_{<H>}> \bullet \nabla \rho^s\} d^3 p - \int_{-\infty}^{+\infty}\int_{-\infty}^{+\infty}\int_{-\infty}^{+\infty} \rho^s \{T\partial_t \Delta S + <\overset{\bullet}{x}_{<H>}> \bullet T\nabla \Delta S\} d^3 p$$

$$= \int_{-\infty}^{+\infty}\int_{-\infty}^{+\infty}\int_{-\infty}^{+\infty} kT\nabla \bullet (\rho^s D\nabla\phi(1 + A + O(\nabla\phi^2))) d^3 p - \int_{-\infty}^{+\infty}\int_{-\infty}^{+\infty}\int_{-\infty}^{+\infty} \rho^s \{T\partial_t \Delta S + <\overset{\bullet}{x}_{<H>}> \bullet T\nabla \Delta S\} d^3 p$$

where it has been used the relation

$$S^s = S + \Delta S$$

and hence that

$$\partial_t n\underline{\mathrm{Y}} + \nabla \bullet n\underline{\mathrm{Y}\overset{\bullet}{q}} - F \bullet \underline{\overset{\bullet}{q}} - \int_{-\infty}^{+\infty}\int_{-\infty}^{+\infty}\int_{-\infty}^{+\infty} \rho^s \{\partial_t \Delta\mathrm{Y} + <\overset{\bullet}{x}_{<H>}> \bullet \nabla(\Delta\mathrm{Y})\} d^3 p$$

$$- \int_{-\infty}^{+\infty}\int_{-\infty}^{+\infty}\int_{-\infty}^{+\infty} \rho^s \{T\partial_t \Delta S + <\overset{\bullet}{x}_{<H>}> \bullet T\nabla \Delta S\} d^3 p$$

$$= - \int_{-\infty}^{+\infty}\int_{-\infty}^{+\infty}\int_{-\infty}^{+\infty} \rho^s \{S\partial_t T + <\overset{\bullet}{x}_{<H>}> \bullet S\nabla T\} d^3 p + \int_{-\infty}^{+\infty}\int_{-\infty}^{+\infty}\int_{-\infty}^{+\infty} kT(\phi - 1)\{\nabla \bullet \rho^s D\nabla\phi(1 + A + O(\nabla\phi^2))\} d^3 p$$

$$\partial_t n\underline{\mathrm{Y}} + \nabla \bullet n\underline{\mathrm{Y}\overset{\bullet}{q}} - F \bullet \underline{\overset{\bullet}{q}} + \Delta_1$$

$$= - \int_{-\infty}^{+\infty}\int_{-\infty}^{+\infty}\int_{-\infty}^{+\infty} \rho^s S\{\frac{dT}{dt}\} d^3 p + \int_{-\infty}^{+\infty}\int_{-\infty}^{+\infty}\int_{-\infty}^{+\infty} kT(\phi - 1)\{\nabla \bullet \rho^s D\nabla\phi(1 + A + O(\nabla\phi^2))\} d^3 p$$

where

$$\Delta_1 = \int_{-\infty}^{+\infty}\int_{-\infty}^{+\infty}\int_{-\infty}^{+\infty} \rho^s \{\partial_t \Delta\mathrm{Y} + <\overset{\bullet}{x}_{<H>}> \bullet \nabla(\Delta\mathrm{Y})\} d^3 p + \int_{-\infty}^{+\infty}\int_{-\infty}^{+\infty}\int_{-\infty}^{+\infty} \rho^s \{T\partial_t \Delta S + <\overset{\bullet}{x}_{<H>}> \bullet T\nabla \Delta S\} d^3 p$$

$$\partial_t n\underline{\mathrm{Y}} + \nabla \bullet n\underline{\mathrm{Y}\overset{\bullet}{q}} - F \bullet \underline{\overset{\bullet}{q}} - \Delta_1$$

$$= -\frac{\alpha}{k} \int_{-\infty}^{+\infty}\int_{-\infty}^{+\infty}\int_{-\infty}^{+\infty} \rho^s S\{\frac{d<E>}{dt}\} d^3 p + \int_{-\infty}^{+\infty}\int_{-\infty}^{+\infty}\int_{-\infty}^{+\infty} kT(\phi - 1)\{\nabla \bullet \rho^s D\nabla\phi(1 + A + O(\nabla\phi^2))\} d^3 p$$

$$\partial_t n\underline{\Upsilon} + \nabla \bullet n\underline{\Upsilon \dot{q}} - F \bullet \underline{\dot{q}} + \frac{\gamma}{k}\int_{-\infty}^{+\infty}\int_{-\infty}^{+\infty}\int_{-\infty}^{+\infty} \rho^s S\{\frac{dE}{dt}\} d^3p$$

$$= \int_{-\infty}^{+\infty}\int_{-\infty}^{+\infty}\int_{-\infty}^{+\infty} kT(\phi - 1)\{\nabla \bullet \rho^s D \nabla \phi (1 + A + O(\nabla \phi^2))\} d^3p + \Delta_1$$

$$\partial_t n\underline{\Upsilon} + \nabla \bullet n\underline{\Upsilon \dot{q}} - F \bullet \underline{\dot{q}} + \frac{\gamma}{k} F \bullet \underline{S\dot{q}}$$

$$= \int_{-\infty}^{+\infty}\int_{-\infty}^{+\infty}\int_{-\infty}^{+\infty} kT(\phi - 1)\{\nabla \bullet \rho^s D \nabla \phi (1 + A + O(\nabla \phi^2))\} d^3p + \Delta_1$$

that can synthetically read

$$\partial_t n\underline{\Upsilon} + \nabla \bullet n\underline{\Upsilon \dot{q}} - F \bullet \underline{\dot{q}} + \frac{\alpha}{k} F \bullet \underline{S\dot{q}}$$

$$= -\int_{-\infty}^{+\infty}\int_{-\infty}^{+\infty}\int_{-\infty}^{+\infty} kT \frac{(\phi - 1)}{\phi} \rho^s (\frac{d_s \phi}{dt}) d^3p + \Delta_0 + \Delta_1$$

where

$$n\underline{S\dot{q}} = \int_{-\infty}^{+\infty}\int_{-\infty}^{+\infty}\int_{-\infty}^{+\infty} \rho^s S < \dot{q} > d^3p$$

$$\Delta_0 = \{\int_{-\infty}^{+\infty}\int_{-\infty}^{+\infty}\int_{-\infty}^{+\infty} \phi^2 (\nabla \bullet \nabla \phi)(1 + A + O(\nabla \phi^2)) d^3p\}$$

Equation (???) is the continuity equation for the generalised SQHA free energy density $\Upsilon$ .
Integrating (???) over a volume V and using the Gauss theorem, we find:

$$\iiint_V \partial_t n\underline{\Upsilon}\, d^3q + \oiint_\Sigma \nabla \bullet n\underline{\Upsilon \dot{q}}\, d\sigma - \iiint_V F \bullet (\underline{\dot{q}} - \frac{\gamma}{k}\underline{S\dot{q}})\, d^3q$$

$$= -\iiint_V \int_{-\infty}^{+\infty}\int_{-\infty}^{+\infty}\int_{-\infty}^{+\infty} kT \rho^s (\frac{d_s \phi}{dt}) d^3p\; d^3q + \iiint_V \Delta_0\; d^3q + \iiint_V \Delta_1\; d^3q$$

If the integration is done on the volume of the system the first term is the total time derivative of the generalised free energy (GFE) such as:

$$\iiint_V \partial_t n\underline{\Upsilon}\, d^3q = \frac{d\Phi}{dt}$$

the second term represents that one that leaves the system due to the molecular flow through the boundary (positive outgoing) that, it results

$$\oiint_{\Sigma} \nabla \bullet n\underline{\Upsilon \dot{q}}\, d\sigma = -\frac{d\Phi_{sup}}{dt}$$

The third term represents the GFE variation due to the volume force of the external reservoirs (particles kinetic energy $dE$ /dt) plus GFE generated by flux of entropi that read

$$\iiint_V F \bullet \underline{\dot{q}}\, d^3 q = \frac{dL_{ext(vol)}}{dt} = \frac{dE_{vol}}{dt} = \frac{dE_{cin}(\text{vol})}{dt} + \frac{dE_{int}(\text{vol})}{dt}$$

$$\frac{\gamma}{k}\iiint_V F \bullet \underline{S\dot{q}}\, d^3 q = \frac{dTS^s{}_{vol}}{dt}$$

Hence, we find at constant volume

$$\frac{d\Phi}{dt} - \frac{d\Phi_{sup}}{dt} - \frac{d(E_{cin} + E_{int})}{dt} + \frac{dTS^s{}_{vol}}{dt} = -\iiint_V \int_{-\infty}^{+\infty}\int_{-\infty}^{+\infty}\int_{-\infty}^{+\infty} kT\frac{(\phi-1)}{\phi}\rho^s\left(\frac{d_s\phi}{dt}\right)d^3p\ d^3q + \Delta_0 + \Delta_1$$